\documentclass[lettersize,journal]{IEEEtran}
\usepackage{amsmath,amsfonts}
\usepackage[ruled,linesnumbered]{algorithm2e}
\usepackage{array}
\usepackage[caption=false,font=normalsize,labelfont=sf,textfont=sf]{subfig}
\usepackage{textcomp}
\usepackage{stfloats}
\usepackage{url}
\usepackage{verbatim}
\usepackage{graphicx}
\usepackage{cite}
\usepackage{hyperref}
\usepackage{tikz}
\usepackage{hwemoji}
\usepackage{amssymb}
\usepackage{threeparttable}
\usepackage{multirow}
\usepackage{pifont}
\usepackage{tabularx}
\usepackage{orcidlink}
\usepackage{makecell}
\usepackage{booktabs}
\begin{document}

\title{FireFly-T: High-Throughput Sparsity Exploitation for Spiking Transformer Acceleration with Dual-Engine Overlay Architecture}

\author{
  Tenglong Li \orcidlink{0009-0007-3266-2075},
  Jindong Li \orcidlink{0000-0002-4009-916X},
  Guobin Shen \orcidlink{0000-0002-4069-2107},
  Dongcheng Zhao \orcidlink{0000-0002-0593-8650}, 
  Qian Zhang \orcidlink{0000-0001-5314-4233},
  Yi Zeng \orcidlink{0000-0002-9595-9091}
  
  \thanks{Received 19 May 2025. This work was supported by the General Program of the National Natural Science Foundation of China (32571284, 62406325, and 32441109), the State Key Laboratory of Brain Cognition and Brain-inspired Intelligence Technology (JS202401), the Artificial Intelligence of General Anesthesia Mechanism Program of Yichang Humanwell (Y8D1051LE1), and the Institute of Automation, Chinese Academy of Sciences (E411230101). \textit{(Corresponding authors: Qian Zhang; Yi Zeng.)}}

  \thanks{Tenglong Li, Jindong Li, and Qian Zhang are with the School of Artificial Intelligence, University of Chinese Academy of Sciences, Beijing 100049, China.}

  \thanks{Guobin Shen is with the School of Future Technology, University of Chinese Academy of Sciences, Beijing 100049, China.}

  \thanks{Qian Zhang, Guobin Shen, Dongcheng Zhao, and Yi Zeng are with the Center for Long-term AI, Beijing 101407, China.}

  \thanks{Yi Zeng is also with the State Key Laboratory of Brain Cognition and Brain-inspired Intelligence Technology, Chinese Academy of Sciences, Shanghai 200031, China.}

  \thanks{All authors are with the Brain-inspired Cognitive AI Lab, Institute of Automation, Chinese Academy of Sciences, Beijing 100190, China. (Emails: litenglong2023@ia.ac.cn; lijindong2022@ia.ac.cn; shenguobin2021@ia.ac.cn; zhaodongcheng2016@ia.ac.cn; q.zhang@ia.ac.cn; yi.zeng@ia.ac.cn)}
}

\markboth{IEEE Transactions on Computers,~Vol.~75, No.~6, June~2026}%
{Li \MakeLowercase{\textit{et al.}}: FireFly-T: High-Throughput Sparsity Exploitation for Spiking Transformer Acceleration}


\maketitle

\begin{abstract}
Spiking transformers are emerging as a promising architecture that combines the energy efficiency of Spiking Neural Networks (SNNs) with the powerful attention mechanisms of transformers.
However, existing hardware accelerators lack support for spiking attention, exhibit limited throughput when exploiting fine-grained sparsity, and struggle with scalable parallelism in sparse computation.
To address these challenges, we propose FireFly-T, a dual-engine overlay architecture that integrates a sparse engine for activation sparsity and a binary engine for spiking attention.
In the sparse engine, we present a high-throughput sparse decoder that exploits fine-grained sparsity by concurrently extracting multiple non-zero spikes. To complement this, we introduce a scalable load balancing mechanism with weight dispatch and out-of-order execution, eliminating bank conflicts to support scalable multidimensional parallelism.
In the binary engine, we leverage the byte-level write capability of SRAMs to efficiently manipulate the 3D dataflows required for spiking attention with minimal resource overhead. We also optimize the core AND-PopCount operation in spiking attention through a LUT6-based implementation, improving timing closure and reducing LUT utilization on Xilinx FPGAs.
As an overlay architecture, FireFly-T further incorporates an orchestrator that dynamically manipulates input dataflows with flexible adaptation to diverse network topologies, while ensuring efficient resource utilization and maintaining high throughput.
Experimental results demonstrate that our accelerator achieves $1.39\times$ and $2.40\times$ higher energy efficiency, as well as $4.21\times$ and $7.10\times$ greater DSP efficiency, compared to FireFly v2 and the transformer-enabled SpikeTA, respectively. These results highlight its potential as an efficient hardware platform for spiking transformers.
\end{abstract}


\vfill\break
\section{Introduction} \label{sec:introduction}

\IEEEPARstart{T}{ransformer} networks have revolutionized domains such as natural language processing~\cite{vaswani2017attention, radford2019language, devlin2019bert} and computer vision~\cite{dosovitskiy2020image, yuan2021tokens, radford2021learning}. The Vision Transformer (ViT), for example, processes images as patch sequences and achieves state-of-the-art results~\cite{dosovitskiy2020image}. Meanwhile, Spiking Neural Networks (SNNs), inspired by the biological brain~\cite{maass1997networks}, are drawing attention for their energy efficiency. This efficiency arises from their binary activations and inherently sparse computation, with studies showing that SNNs naturally exhibit high sparsity without the need for explicit pruning~\cite{kim2023snpu, yin2024mint}, marking them as promising candidates for energy-efficient real-world applications.

Integrating these two paradigms, spiking transformers combine the powerful attention mechanism of transformers with the efficiency of SNNs, demonstrating competitive performance while reducing computational overhead and power consumption~\cite{zhou2022spikformer, zhou2023spikingformer,yao2023spike, yao2024spikedriven, zhou2024qkformer, shi2024spikingresformer, shen2024conventional}. Further enhancements, such as binary attention mechanisms~\cite{shen2024enhancing, cao2025binary}, offer additional advantages by reducing the bit-width of attention maps and potentially enhancing accuracy, solidifying spiking transformers as an attractive direction for practical deployment.

To fully unlock the potential of spiking transformers, the development of specialized hardware support is crucial—especially for accelerating spiking attention and effectively exploiting activation sparsity. While recent efforts have advanced SNN acceleration, existing solutions still exhibit limitations, falling short when applied to spiking transformers. Specifically, we identify three critical limitations in current hardware approaches, as detailed below:

\begin{figure*}
  \centering
  \includegraphics[width=1.0\linewidth]{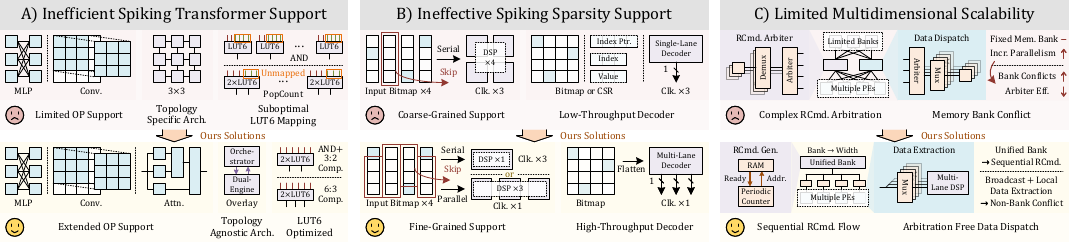}
  \caption{
    Overview of major limitations in hardware acceleration for spiking transformers, alongside the proposed solutions and features of our FireFly-T.
  }
  \label{fig:challenge}
\end{figure*}

\textbf{Inefficient Spiking Transformer Support.} Existing SNN accelerators primarily focus on optimizing multilayer perceptron (MLP) and convolutional layers~\cite{liu2023low, chen2024sibrain, mao2024stellar, yin2024loas, liu2024exploiting}, often overlooking the attention mechanism crucial to spiking transformers. Mapping attention operations onto existing hardware is challenging due to the unique dataflow patterns of the multi-head attention mechanism and inefficiencies arising from bit-width mismatches—spiking attention performs operations between 1-bit operands, whereas accelerators are optimized for spiking convolutions with 1-bit activations and multi-bit weights. Dedicated hardware designs for spiking attention also exhibit limited efficiency. As illustrated in Fig.~\ref{fig:challenge}A, Gao et al.~\cite{gao2025advancing} utilize six-input Look-Up Tables (LUT6) with only two active inputs for AND followed by Population Count (AND-PopCount) operations central to binary attention, resulting in suboptimal resource utilization. Fang et al.~\cite{fang2024energy} adopt a $3\times3$ core organization, which suffers from performance degradation across diverse network topologies, and also inefficiently accumulates 1-bit spikes through an 8-bit adder tree.

\textbf{Ineffective Spiking Sparsity Support.} Achieving high throughput while effectively exploiting spiking sparsity remains a major challenge. Several prior studies have attempted to exploit coarse-grained sparsity~\cite{yin2024loas,fang2024energy}, which naturally leads to more regular dataflow and improved workload balance relative to fine-grained methods. However, this approach often results in suboptimal sparsity utilization. As depicted in Fig.~\ref{fig:challenge}B, under 75\% uniform sparsity, grouping spikes in fours yields only a $\sim$31\% chance of skipping an entire group. This limits performance gains, while still requiring four DSPs per group, some of which remain idle during processing. On the other hand, while fine-grained methods are more effective at exploiting sparsity, they often suffer from low input processing throughput. Existing sparse accelerators that decode Compressed Sparse Row (CSR) or bitmap formats typically process only one non-zero activation per processing element (PE) per cycle~\cite{liu2024exploiting, gondimalla2019sparten}, leading to lower input bandwidth utilization than dense or coarse-grained approaches~\cite{li2024firefly}.

\textbf{Limited Multidimensional Scalability.} 
Dense accelerators—such as DeepFire2~\cite{aung2023deepfire2} and FireFly v2~\cite{li2024firefly}—benefit from regular dataflows that naturally enable extensive parallelism across multiple dimensions, including spatial, temporal, and channel dimensions in SNNs. However, this regularity comes at the cost of reduced energy efficiency when processing sparse inputs and limited inherent support for sparsity. In contrast, sparse accelerators face challenges in achieving comparable multidimensional parallelism. Their irregular dataflows and the resulting workload imbalances often limit parallelism to a narrower set of dimensions~\cite{yin2024loas, liu2024exploiting}. A primary contributing factor is the overhead introduced by sparsity management mechanisms. As shown in Fig.~\ref{fig:challenge}C, workload balancing techniques using crossbars~\cite{yin2024loas, yang2024trapezoid} can introduce severe memory bank conflicts caused by weight reuse—especially when scaling pixel- or temporal-level parallelism—which degrade throughput and limit overall performance scaling.

In this work, we present \textit{\textbf{FireFly-T}}, an architecture designed to accelerate both spiking convolutional networks and spiking transformers. Extending our FireFly series, the \textit{T} denotes enhanced transformer support. FireFly-T addresses the challenges outlined in Fig.~\ref{fig:challenge} with its solutions depicted in the lower part. Our main contributions are summarized below:

\begin{itemize}
  \item We propose a dual-engine overlay architecture featuring dedicated sparse and binary engines coupled with flexible orchestrator. This design enables efficient support for spiking transformers with latency-hiding pipelines and multidimensional parallelism, while sustaining high performance across diverse network topologies.
  \item We present a high-throughput sparse engine capable of effective fine-grained sparsity exploitation, incorporating multi-lane sparse decoders and a bank-conflict-free load balancing mechanism. This contributes to comparable throughput while reducing DSP usage by 75\% compared to FireFly v2, improving overall energy efficiency.
  \item We implement a resource-efficient binary engine featuring an implicit transpose mechanism, combined with LUT6-optimized AND-PopCount logic. This design eliminates the need for additional transposition buffers and reduces LUT utilization by 52\%, enhancing overall resource efficiency for spiking attention support.
\end{itemize}

The remainder of this paper is organized as follows: Section~\ref{sec:bg} provides background on spiking attention and the structure of the spiking transformer. Section~\ref{sec:basic} introduces the fundamentals of FireFly-T. Section~\ref{sec:arch} presents our proposed accelerator microarchitecture, including the sparse engine, binary engine, and orchestrator implementation. Section~\ref{sec:experiments} describes the experimental setup and provides a comprehensive evaluation. Section~\ref{sec:related} reviews related work, and Section~\ref{sec:conclusion} concludes the paper and outlines directions for future research.

\section{Background} \label{sec:bg}
\subsection{Spiking Attention Mechanism} \label{subsec:attention}
The attention mechanism is a core component of transformer architectures, generating context-aware representations by weighting value vectors ($V$) based on the attention scores computed from queries ($Q$) and keys ($K$)~\cite{vaswani2017attention, radford2019language}. In spiking transformer models~\cite{zhou2022spikformer, zhou2023spikingformer}, this mechanism is adapted to work with spiking neural dynamics ($SN$), typically using Leaky Integrate-and-Fire (LIF) neurons~\cite{dayan2005theoretical} to introduce nonlinearity and improve energy efficiency. Given an input sequence $X \in \mathbb{R}^{L \times d}$, with sequence length $L$ and embedding dimension $d$, the $Q$, $K$, and $V$ matrices are obtained via linear projections with learnable weight matrices $W_Q, W_K, W_V \in \mathbb{R}^{d \times d}$, followed by spiking neuron activation:
\begin{equation}
    Q = SN(XW_Q),\ K = SN(XW_K),\ V = SN(XW_V)
\end{equation}

The attention scores are then computed as the dot-product of $Q$ and $K$ scaled by $\sqrt{d}$. Unlike ViT~\cite{dosovitskiy2020image}, which applies a softmax function to normalize the scores, spiking attention~\cite{zhou2022spikformer, zhou2023spikingformer} directly aggregates the result with the value matrix $V$:
\begin{equation}
    \text{Attention}(Q, K, V) = SN(\frac{QK^T}{\sqrt{d}} V)
\end{equation}

In pursuit of greater computational efficiency, as shown in Table~\ref{tab:network}, recent studies have explored binary attention mechanisms~\cite{cao2025binary, shen2024enhancing, yao2023spike, yao2024spikedriven}, which transform both the attention scores $QK^T$ and the weighted outputs $QK^TV$ into spiking form. Two prevalent approaches are binarization via spiking neurons and thresholding. The former employs spiking neurons, while the latter emits a spike only if the input current exceeds a learnable threshold $\Delta$. Beyond reducing computational overhead, this binarization enhances network performance potentially due to the induced non-linearity. For instance, under the same CIFAR-100 setting, \cite{shen2024enhancing} achieves a 0.95\% accuracy improvement compared to Spikformer~\cite{zhou2022spikformer} despite the lower compute cost.

\begin{figure}
  \centering
  \includegraphics[width=1.0\linewidth]{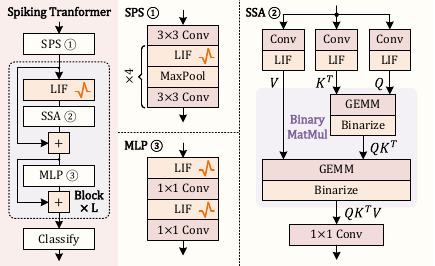}
  \caption{
    Overview of typical spiking transformer~\cite{shen2024enhancing, cao2025binary}, which begins with a SPS module, followed by $L$ encoder blocks and a classification head. Each encoder block includes a SSA, an MLP, and associated residual connections.
  }
  \label{fig:spikingformer}
\end{figure}

\begin{table}
  \setlength{\tabcolsep}{3.1pt}
  \centering
  \renewcommand\arraystretch{1.25}
  \begin{threeparttable}[b]
  \caption{Comparison of typical spiking transformer architectures}
  \begin{tabular}{ccccc}
  \toprule
  \multirow{2}{*}{\textbf{Work}} & \textbf{Binary} & \textbf{Residual} & \textbf{Network} & \textbf{Compute}\\
   & \textbf{Attention} & \textbf{Connection} & \textbf{Accuracy} & \textbf{Efficiency}\\
  \midrule
  Spikformer~\cite{zhou2022spikformer} & No & Post-Neuron & Low & Low \\
  Spikingformer~\cite{zhou2023spikingformer} & No & Pre-Neuron & Medium & Medium \\
  Shen et al.~\cite{shen2024enhancing} & Yes & Post-Neuron & Medium & Medium \\
  SDT~\cite{yao2023spike} \& SDTv2~\cite{yao2024spikedriven} & Yes & Pre-Neuron & High & High \\
  BESTformer~\cite{cao2025binary} & Yes & Pre-Neuron & High & High \\
  \bottomrule
  \end{tabular}
  \label{tab:network}
  \end{threeparttable}
\end{table}

\subsection{Spiking Transformer Architecture}
Spiking transformers extend standard transformer architecture to accommodate processing under SNN paradigm~\cite{zhou2022spikformer, zhou2023spikingformer}. A typical architecture consists of three main components: a Spiking Patch Splitting (SPS) module, a stack of encoder blocks, and a classification head, as illustrated in Fig.~\ref{fig:spikingformer}.

The SPS module functions as a hierarchical tokenizer, employing a series of $3\times3$ spiking convolutional layers to progressively downsample the input image while increasing channel dimensions. This process generates compact spike-based patch embeddings, which are then forwarded to the encoder blocks. Each encoder block consists of a Spiking Self-Attention (SSA) module and an MLP. Specifically, the SSA module combines the binary spiking attention mechanism (Section~\ref{subsec:attention}) with a $3\times3$ convolutional layer. The subsequent MLP executes channel expansion and reduction via two $1\times1$ convolutional layers interspersed with spiking neurons.

As summarized in Table~\ref{tab:network}, existing spiking transformers adopt varying strategies for residual connections. Architectures such as Spikformer~\cite{zhou2022spikformer} and Shen et al.~\cite{shen2024enhancing} employ post-neuron residuals, where spikes are added after the spiking neuron. Consequently, subsequent operations (e.g., convolutions) process non-binary inputs, reducing computational efficiency. In contrast, models like Spikingformer~\cite{zhou2023spikingformer}, SDT~\cite{yao2023spike}, and BESTformer~\cite{cao2025binary} adopt pre-neuron residuals. By integrating residuals into the membrane potentials before the spiking neuron (Fig.~\ref{fig:spikingformer}), they retain richer information while keeping downstream operations spike-based. This strategy proves advantageous; for instance, Spikingformer surpasses Spikformer by 2.21\% on ImageNet. Accordingly, we focus on supporting binary attention and pre-neuron residuals, which achieve a favorable balance between performance and hardware efficiency.
\section{FireFly-T Basics} \label{sec:basic}
\subsection{Overall Architecture} \label{subsec:overall}

\begin{figure}
  \centering
  \includegraphics[width=1.0\linewidth]{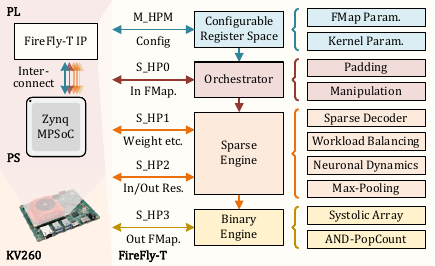}
  \caption{
    Block diagram of the FireFly-T system architecture on a Zynq UltraScale+ MPSoC (KV260), highlighting key features and components of FireFly-T IP, alongside the PS-PL interface for control and data transfer.
  }
  \label{fig:diagram}
\end{figure}

The FireFly-T system is built upon the Processing System (PS) and Programmable Logic (PL) of the Zynq UltraScale+ SoC, as shown in Fig.~\ref{fig:diagram}. For command and status control, an M-AXI-HPM master port on the PS side directly interfaces with the FireFly-T IP. To facilitate high-throughput data transfer between the PS and PL, four 128-bit S-AXI-HP ports are enabled. These ports handle specific tasks: loading weights, biases, and membrane thresholds; fetching feature map inputs; writing feature map outputs; and providing a read-and-write interface for residual input and output data.

Within the FireFly-T IP, the architecture is implemented as an overlay consisting of three main modules: the orchestrator, the sparse engine, and the binary engine. During inference, the feature maps of each network layer are processed in a pipelined manner—first by the orchestrator, then by the sparse engine, and finally by the binary engine if attention is enabled. This processing sequence is applied uniformly across all layers, reusing the same hardware throughout inference.

\begin{figure*}
  \centering
  \includegraphics[width=1.0\linewidth]{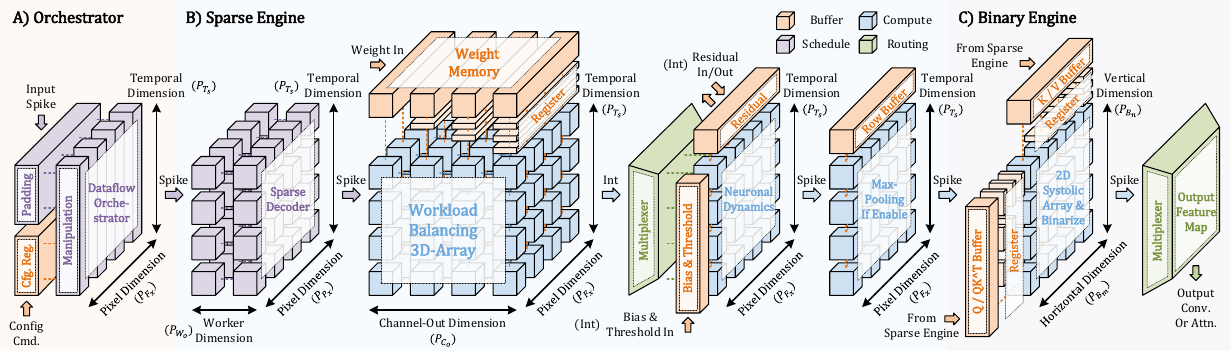}
  \caption{
    Illustration of FireFly-T's parallelism and dataflow, featuring the orchestrator, sparse engine, and binary engine.
  }
  \label{fig:architecture}
\end{figure*}

\textbf{Orchestrator.} The orchestrator functions as the central controller, retrieving layer-specific parameters from the PS and propagating them to downstream engines. It also manages padding and performs dataflow manipulation to prepare inputs for the sparse engine. The data formats span multiple dimensions (e.g., pixel, spatial, channel), allowing the computation engine to effectively exploit multidimensional parallelism.

\textbf{Sparse Engine.} The sparse engine accelerates convolutional and linear operations by exploiting fine-grained activation sparsity. It uses high-throughput multi-lane decoders to extract multiple non-zero indices concurrently, reusing them across output channels for efficiency. The workload-balanced 3D array architecture decouples load balancing from memory banking, supporting spatial, temporal, and channel-in/out parallelism. Pipelined accumulation enables group-wise membrane updates, after which outputs pass through max-pooling and, if attention is enabled, are forwarded to the binary engine.

\textbf{Binary Engine.} The binary engine accelerates binary attention ($QK^T$ and $QK^TV$) using a resource-efficient 2D systolic array. Its AND-PopCount logic is finely optimized to the granularity of LUT6 resources, enhancing both hardware efficiency and timing closure. Leveraging SRAM byte-write capabilities, it performs spiking-attention dataflow permutations without explicit transposition buffers. Closely integrated with the sparse engine, it operates in a pipelined manner, reducing off-chip memory traffic and effectively hiding multi-head attention latency, which will be detailed in Section~\ref{subsec:pipeline}.

\subsection{Parallelism and Dataflow} \label{subsec:parallelism}
Parallelism in SNNs can typically be exploited along four primary dimensions: temporal, spatial (or pixel-wise), input channels, and output channels. FireFly-T is designed to support parallelism across all these dimensions. For reference, key symbols are summarized in Table~\ref{tab:notation}.

\begin{table}
  \centering
  \setlength{\tabcolsep}{1.5pt}
  \renewcommand\arraystretch{1.25}
  \caption{Summary of Key Notations in FireFly-T}
  \label{tab:notation}
  \begin{tabular}{ccc}
  \toprule
  \textbf{Type} & \textbf{Symbol} & \textbf{Description} \\
  \midrule
  \multirow{3}{*}{\makecell{Network \\ Parameter}} & $T_s, L, d$ & Timesteps, Seq. len., Embed. dim. \\
  & $K_{h}, K_{w}; F_{h}, F_{w}$ & Kernel size; Feature map size \\
  & $C_{i}, C_{o}$ & Input channels, Output channels \\
  \midrule
  \multirow{2}{*}{\makecell{Hardware \\ Parallelism}} & $P_{T_s}, P_{F_x}; P_{C_{i}}, P_{C_{o}}$ & Temporal, Spatial, Channel parallelism \\
  & $P_{B_m}, P_{B_n}, P_{B_k}$ & Binary engine parallelism \\
  \midrule
  \multirow{2}{*}{\makecell{Hardware \\ Configuration}} & $M, P_{W_o}$ & Decoder lanes, Balancing workers \\
  & $B_m, G$ & Memory banks, Decoder Throughput \\
  \bottomrule
  \end{tabular}
\end{table}

As illustrated in Fig.~\ref{fig:architecture}A, the orchestrator consists of $P_{T_s}$ memory banks and manages dataflow with a unified format $(P_{T_s}, P_{F_x}, P_{C_i})$, ensuring matched throughput between inputs and outputs. The packed spikes are then routed to a grid of sparse decoders organized along worker and spatiotemporal dimensions (Fig.~\ref{fig:architecture}B). Each decoder consumes a $P_{C_i}$-length spike vector per cycle and its output indices are reused across output channels in the downstream sparse engine.

The core computation is distributed between the sparse and binary engines. The sparse engine employs a 3D workload balancing array structured as $(P_{T_s}, P_{F_x}, P_{C_o})$. By pipelining computation across output channels, the neuronal dynamics module is effectively reduced to a 2D grid, improving both efficiency and scalability. For spiking attention, computation is offloaded to the binary engine, which is organized as a $(P_{B_m} \times P_{B_n})$ systolic array (Fig.~\ref{fig:architecture}C). Each processing element performs AND-PopCount operations on vectors of length $P_{B_k}$. Finally, outputs from both engines are reformatted to match the unified layout required by the overlay architecture, ensuring seamless dataflow between layers.

\subsection{Latency-Hiding Pipeline} \label{subsec:pipeline}
\begin{figure}
  \centering
  \includegraphics[width=1.0\linewidth]{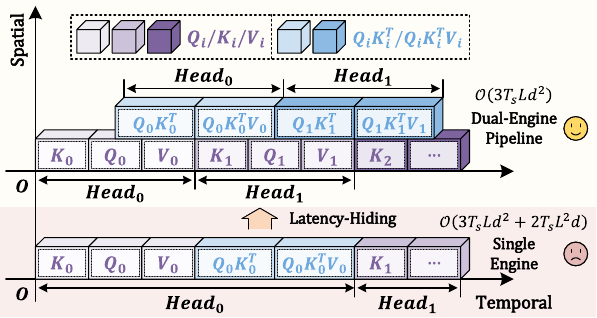}
  \caption{
    Spatiotemporal diagram of the latency-hiding pipeline. The execution of spiking attention is overlapped between the sparse and binary engines.
  }
  \label{fig:pipeline}
\end{figure}

Fig.~\ref{fig:pipeline} illustrates the proposed latency-hiding pipeline diagram, where the binary engine computes attention scores ($QK^T$) and weighted sums ($QK^TV$) concurrently with the sparse engine's projection operations ($Q, K, V$). By overlapping these operations across attention heads, the latency overhead of attention computation can be effectively hidden.

To ensure efficient pipelining without stalling, the throughput of the sparse and binary engines should be well balanced. For per-head latency modeling, we define the computational workloads ($W_s$, $W_b$) and the available parallelism ($P_s$, $P_b$) of the sparse and binary engines as follows:
\begin{equation}
  \begin{aligned}
    W_s &= T_s \times L \times d \times d \\
    W_b &= T_s \times L \times d \times L \\
    P_s &= P_{T_s} \times P_{F_x} \times P_{C_i} \times P_{C_o} \\
    P_b &= P_{B_m} \times P_{B_n} \times P_{B_k}
  \end{aligned}
  \label{eq:pipeline}
\end{equation}

Given that the sparse engine processes three projection phases while the binary engine handles two attention phases, the condition for load balancing is expressed as $3 \times (W_s / P_s) \approx 2 \times (W_b / P_b)$. Substituting the definitions from Eq.~\ref{eq:pipeline} yields the required relationship between the parallelisms:
\begin{equation}
 P_b \approx 2/3 \times (L / d) \times P_s
\end{equation}

In spiking transformer configurations~\cite{shen2024enhancing, zhou2023spikingformer}, the ratio $L/d$ implies that the required $P_b$ is much smaller than $P_s$, ranging approximately from $1/8$ to $1/4$. By further leveraging the 4-lane accumulation capability of the DSP48E2 units, the binary engine efficiently supports spiking attention with minimal hardware resources, while reducing the overall execution time complexity from $\mathcal{O}(3T_sLd^2 + 2T_sL^2d)$ to $\mathcal{O}(3T_sLd^2)$.

\section{FireFly-T Microarchitecture} \label{sec:arch}
\subsection{High-Throughput Sparse Engine}
As outlined in Section~\ref{subsec:overall}, the sparse engine is designed to efficiently exploit the fine-grained sparsity of spiking activations. This is achieved through a multi-lane sparse decoder that supports high-throughput processing of vectorized inputs. In addition, a scalable workload balancing mechanism is introduced, which evenly distributes workloads while avoiding bank conflicts induced by crossbar methods. This approach enables efficient scaling of parallelism across both spatial and temporal dimensions, laying the foundation for the system's multidimensional parallel processing capabilities.

\subsubsection{\textbf{Multi-Lane Sparse Decoder}} \label{subsubsec:decoder}

\begin{figure*}
  \centering
  \includegraphics[width=1.0\linewidth]{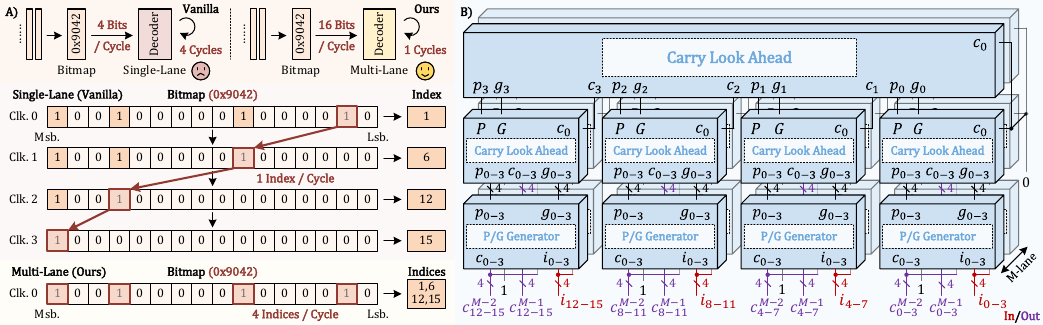}
  \caption{
    Bitmap-based sparse decoder. A1) Throughput comparison between single-lane decoder~\cite{fang2024energy,li2024fireflys} and multi-lane ($M=4$) decoder, using example input 0x9042. A cycle-level decoding process is shown for both single-lane and multi-lane ($M=4$) decoder. B) Architecture of a multi-lane ($M=2$) sparse decoder, detailing its intra- and inter-block grouped carry generator organization.
  }
  \label{fig:decoder}
\end{figure*}

We identified a performance bottleneck in existing sparsity-oriented SNN accelerators due to their limited ability to exploit input throughput, which typically corresponds to channel-in parallelism. This stems from the conventional sparse decoder designs: bitmap-based decoders produce only one output (a one-hot vector or index) per cycle, and CSR-based decoding requires serial traversal of column indices~\cite{fang2024energy, parashar2017scnn, wang2024compass}. As illustrated in Fig.~\ref{fig:decoder}A, with 16-bit input bandwidth, decoding the bitmap input 0x9042 on a single-lane decoder takes four cycles, resulting in an effective throughput of just 1/4, and thus underutilizing the available channel-in parallelism. 

To address this limitation, we present a multi-lane sparse decoder with $M$ lanes, capable of identifying multiple non-zero spike indices in parallel from a sparse
input vector of length \(N\), where \(N = P_{C_i}\). As depicted in Fig.~\ref{fig:decoder}A, a 4-lane decoder ($M = 4$) can process the input bitmap 0x9042 in a single cycle by simultaneously extracting four active spikes, thereby improving throughput. To achieve this, our design draws inspiration from lookahead carry generators, where each bit position uses propagate ($p$) and generate ($g$) signals to compute the $(n+1)$-th carry-out $c_{n+1}$ as $c_{n+1} = g_n \lor (p_n \land c_n)$. We adapt this principle to the architecture of our $M$-lane sparse decoder. Specifically, given a sparse input $i$, the $m$-th lane of the one-hot output $o$ is computed as follows:
\begin{equation}
  \begin{aligned}
    o_n^m &= (i_n \land c_n^{m-1}) \land \neg c_n^m \\
    &= g_n^m \land \neg c_n^m \\
    c_{n+1}^m &= (i_n \land c_n^{m-1}) \lor c_n^m \\
    &= g_n^m \lor (p_n^m \land c_n^m) \\
    \text{where } g_n^m = i_n\, \land&\, c_n^{m-1} \text{, } p_n^m = c_n^{-1} = 1 \text{ and }c_0^m = 0
  \end{aligned}
\end{equation}

Here, $p_n^m$ is always set, and $g_n^m$ indicates that the $n$-th input bit and the carry-in from lane $(m-1)$ are both active, where $c_n^{m-1}$ signifies that the $(m-1)$-th spike has been found prior to position $n$. The output $o_n^m$ is activated if $g_n^m$ is set—meaning the $n$-th input bit is a candidate for the $m$-th spike—and no preceding bit has already triggered the current lane ($\neg c_n^m$).

To ensure correct operation across cycles, the input bitmap is updated as $i_n \leftarrow i_n \land c_{n+1}^{M-1}$, which clears bits that have already been processed. Meanwhile, an input tracker—initialized by a pop-count of the original input and decremented by the throughput $M$ per cycle—controls new input loading, permitting it when its value is less than or equal to $M$.

This lookahead-based architecture offers distinct advantages over conventional approaches. One alternative is to buffer $M$ consecutive inputs, each processed by an independent priority encoder; however, this incurs $M\times$ register overhead. Another approach cascades $M$ priority encoders with iterative masking, which results in a linear increase in logic delay. In contrast, our multi-lane design leverages the lookahead carry principle to extract multiple spikes concurrently from a single input, thus achieving the equivalent throughput with lower buffering overhead and a reduced critical path delay.

Furthermore, to guarantee scalability for wide inputs, the decoder adopts a hierarchical two-level grouped structure, as shown in Fig.~\ref{fig:decoder}B. At the intra-block level, the $N$-bit input is partitioned into $N/k$ blocks of $k$ bits each, each independently computing local propagate and generate signals, defined as
\begin{equation}
  \begin{aligned}
    P^m = \bigwedge_{j=0}^{k-1} p_j^m \text{ and }
    G^m = \bigvee_{j=0}^{k-1} ( g_j^m \land \bigwedge_{l=j+1}^{k-1} p_l^m )
  \end{aligned}
\end{equation}
At the inter-block level, these block-level signals are combined by a secondary lookahead unit to resolve carries across blocks. This hierarchical organization can help reduce fan-in requirements and regularize the layout, thereby mitigating the practical delay penalties associated with wide single-level implementations. By integrating these features, our sparse decoder achieves high throughput and efficient channel-in parallelism within a compact hardware footprint.

\subsubsection{\textbf{Scalable Load Balancing}} \label{subsubsec:balancing}

\begin{figure*}
  \centering
  \includegraphics[width=1.0\linewidth]{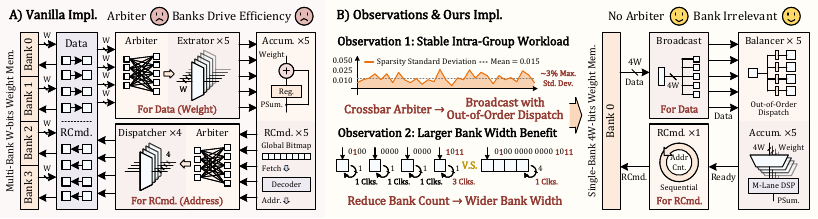}
  \caption{
    Workload balancing approaches. A) Vanilla crossbar-based method~\cite{yin2024loas} with four $W$-bit weight memory banks. B) Our proposed method motivated by the observations, with a single $4W$-bit weight memory bank. Both are demonstrated with the configuration $P_{C_o} = 1$, $P_{T_s} \times P_{F_x} = 5$ and $P_{W_o} = 4$.
  }
  \label{fig:balancing}
\end{figure*}

As discussed in Section~\ref{sec:introduction}, existing SNN sparse accelerators either lack support for multidimensional parallelism~\cite{liu2024exploiting, panchapakesan2022syncnn}, or rely on crossbar-based workload balancing~\cite{yin2024loas, parashar2017scnn}. However, such crossbar-based interconnects typically suffer from performance degradation when scaling parallelism due to increased bank conflicts.

Fig.~\ref{fig:balancing}A shows a crossbar-based workload balancing approach employing a banked memory architecture with an all-to-all interconnect~\cite{yin2024loas}. In this design, kernel weights' input channels are partitioned into $W$-bit chunks and distributed across four memory banks. An initial arbiter is required to manage concurrent data dispatch from these banks to PEs. Upon receiving a data chunk, the extractor within each PE selects the valid weights based on the non-zero spikes identified by the sparse decoder for membrane potential accumulation.

Subsequently, for fetching further valid weights, each PE decodes its global bitmap to locate the address of the next relevant chunk and issues read commands. A second arbiter then resolves potential access conflicts among these commands before they are dispatched to the memory banks. This dual-arbiter structure introduces additional logic overhead and degrades performance due to increased latency from access conflict resolution, particularly when the number of PEs does not match the number of banks. For instance, with five PEs and four banks, at least one PE will inevitably stall—unless it accesses the same address and the same bank as another PE. 

To set the stage for our proposed architecture, we first discuss two key observations that informed our approach:

\textbf{Observation 1.} The sparsity within a kernel window of size $K_h \times K_w \times C_i$ remains relatively stable across the $P_{T_s} \times P_{F_x}$ dimensions, as empirically demonstrated in Fig.~\ref{fig:balancing}B, where the results show that the observed standard deviation is approximately 3\% of the theoretical maximum. This is intuitively plausible given the typically large kernel size—particularly in transformer-based architectures—and the inherent spatiotemporal locality of the $P_{T_s} \times P_{F_x}$ input space.

\textbf{Observation 2.} Processing larger vectorized inputs tends to offer higher performance compared to handling multiple smaller vectors. As depicted in Fig.~\ref{fig:balancing}B, consider four PEs, each with a sparse decoder throughput of 1, receiving input vectors of 0x4, 0x0, 0x0, and 0xB, respectively. In this configuration, the last PE would require three cycles to complete its task. By contrast, a single PE with a decoder throughput of 4 can process the combined input 0x400B in just one cycle, demonstrating enhanced computational throughput.
\begin{figure}
  \centering
  \includegraphics[width=1.0\linewidth]{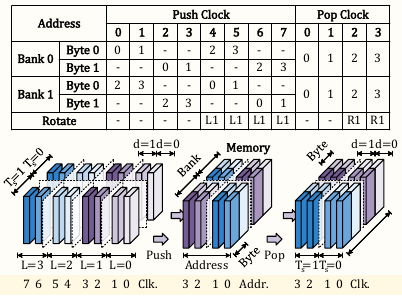}
  \caption{
    Cycle-level illustration of implicit dataflow manipulation transforming the data layout from $(L, d, T_s)$ to $(T_s, d, L)$, with $L = 4$ and $d = T_s = 2$. L1 and R1 indicate one-bank left and right rotations across SRAM banks.
  }
  \label{fig:transpose}
\end{figure}

Leveraging these observations, we propose a unified memory architecture. As shown in the right part of Fig.~\ref{fig:balancing}B, the architecture features a single memory bank that directly broadcasts wider vectorized data across the $P_{T_s} \times P_{F_x}$ dimension. To scale parallelism along the output channel dimension ($P_{C_o}$), the architecture is instantiated per channel, while the indices from sparse decoders are reused. This leads to a 3D processing array organized as $(P_{C_o}, P_{T_s}, P_{F_x})$. In this array, we incorporate an additional worker dimension, $P_{W_o}$, enabling out-of-order dispatch of computations across the $K_h \times K_w \times C_i$ dimensions to idle workers, improving workload balance. Each worker is responsible for extracting valid weight data corresponding to $M$ non-zero indices—provided by the $M$-lane sparse decoder—for membrane potential accumulation.

This design directly leverages \textbf{Observation 1}: stable intra-group sparsity ensures the efficiency of broadcasting a large, vectorized data chunk from the unified, wide memory bank, as most PEs can retrieve their required weights within the broadcast. The wide-bank structure also aligns with the wider vector operations discussed in \textbf{Observation 2}. Integrated with our multi-lane sparse decoders, this enables efficient extraction of multiple weights from the wide vector, enhancing computational throughput. Unlike crossbar-based designs—which require more memory banks to scale with spatiotemporal parallelism—our approach eliminates the need for arbiters and multi-banks, achieving more stable performance scaling under fixed vector lengths, as detailed in Section~\ref{subsec:balancing}.
\begin{figure}
  \centering
  \includegraphics[width=1.0\linewidth]{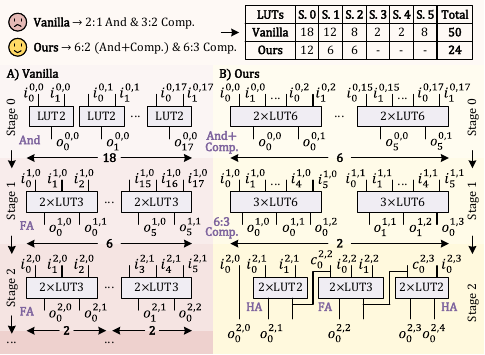}
  \caption{
    Comparison between the vanilla AND-PopCount~\cite{gao2025advancing} and our LUT6-optimized approach with two 18-bit input vectors. Here, $i_k^{m,n}$ denotes the $n$-th bit of the $k$-th input at stage $m$, propagated from output $o_k^{m-1,n}$.
  }
  \label{fig:popcount}
\end{figure}

\subsection{Resource-Efficient Binary Engine}
As described in Section~\ref{subsec:overall}, the binary engine enables efficient binary attention through latency hiding and minimal resource overhead. This section details the key techniques that underpin its resource efficiency, including implicit dataflow manipulation and LUT6-optimized AND-PopCount logic.

\subsubsection{\textbf{Implicit Dataflow Manipulation}}
Consider matrices $Q$, $K$, and $V$ at a specific timestep, each of size $L \times d$, where $L$ denotes the sequence length and $d$ the embedding dimension. When computing $QK^T$, with $Q$ and $K$ stored in row-major order, the inner products are computed row-by-row, producing an $L \times L$ attention score. However, when multiplying the result with $V$, the inner products are computed row-by-column. Hence, to ensure consistent row-wise memory access and reuse the read logic, $V$ must be transposed to have dimensions $d \times L$.

To avoid explicit transposition buffers and repeated instantiation of the 2D systolic array caused by the innermost temporal mapping, we propose an implicit dataflow manipulation technique that treats the SRAM as a three-dimensional structure—(bank, address, byte)—and leverages its byte-write granularity to perform in-place data manipulation during push operations, thereby eliminating the need for an explicit buffer.

Fig.~\ref{fig:transpose} presents a cycle-level example of the proposed implicit dataflow manipulation on the value matrix $V$, reordering the data layout from $(L, d, T_s)$ to $(T_s, d, L)$. During the push phase, data is written into $T_s$ banked RAMs using interleaved addressing with a stride of $d$. In the pop phase, consistent addressing across banks allows permutation of the $d$ and $T_s$ dimensions. Within each bank, $L \times d$ data is streamed in, with push addresses incrementing cyclically along the $d$ dimension and the byte-write mask cyclically rotates along the $L$ dimension, thereby introducing an additional permutation across the $d$ and $L$ dimensions. To fully cover the interleaved address space, both push addresses and input data are rotated across banks every $L$ push cycles. This rotation is reversed during the pop phase to preserve alignment and ensure the correct transposed layout is reconstructed.

Through this overall process, the 3D matrix layout is implicitly manipulated without the need for additional buffering. Such an approach not only minimizes hardware overhead but also enables flexible alignment of the sparse engine's output throughput with the binary engine's input bandwidth, supporting the requirements of the latency-hiding pipeline.

\subsubsection{\textbf{LUT6-Optimized AND-PopCount Logic}} In binary attention mechanisms, the inner product computation within the PEs of systolic array reduces to a bitwise AND followed by a population count. As this constitutes a fundamental computation, optimizing the AND-PopCount operation is crucial for improving the resource efficiency of the binary engine.

\begin{figure*}
  \centering
  \includegraphics[width=1.0\linewidth]{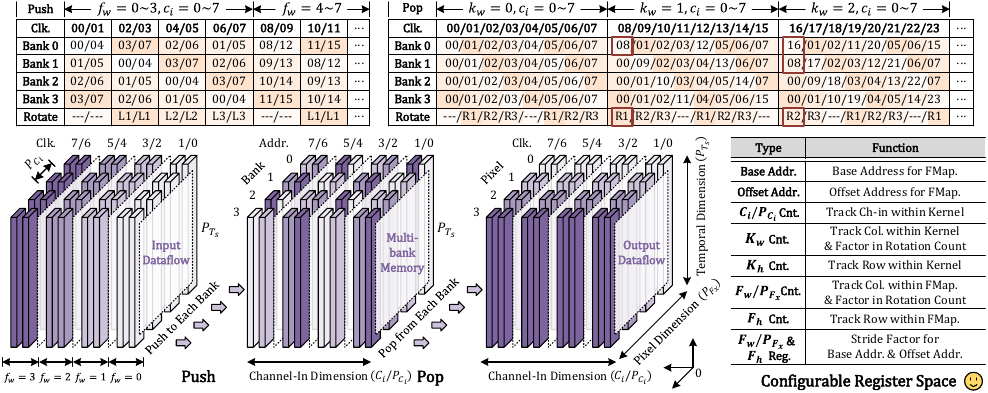}
  \caption{
    Cycle-level illustration of the orchestrator's push and pop phases, demonstrating the interleaved addressing mechanism for data layout transformation. The example uses $F_w = C_i = 8$, $K_h = K_w = 3$, and $P_{F_x} = 4$. Parameters such as $C_o$, $F_h$, $P_{T_s}$, and $P_{C_i}$ are omitted here for simplicity.
  }
  \label{fig:orchestrator}
\end{figure*}

Xilinx FPGAs are built around LUT6 logic units, each capable of implementing arbitrary boolean functions with up to six inputs and one output. However, the straightforward implementation of the AND-PopCount as used by Gao et al.~\cite{gao2025advancing} does not fully utilize the capacity of these LUT6s. As shown in Fig.~\ref{fig:popcount}A, AND gates typically require only two inputs, and directly implementing the popcount logic by Full Adders (FAs, or 3:2 compressors) and Half Adders (HAs) also fails to exploit the full six-input granularity of the LUT6. As a result, such designs lead to inefficient resource utilization and increased logic depth, which worsens timing performance.

To address these inefficiencies, we propose an optimized AND-PopCount implementation that configures LUT6s as specialized compressors. As illustrated in Fig.~\ref{fig:popcount}B, in the initial stage, two LUT6s function as a 6:2 compressor, performing both the bitwise AND operations and an initial population count simultaneously. Subsequent stages employ efficient 6:3 compressors, constructed from three LUT6s, to further reduce the intermediate sums. The final two operands are then processed using a carry-propagating adder composed of FAs and HAs, yielding the final population count.

This multi-stage reduction architecture~\cite{wallace1964suggestion} with dedicated compressors takes full advantage of the six-input granularity of LUT6s. Compared to the naive approach~\cite{gao2025advancing}, it significantly reduces both logic depth and LUT6 consumption. For example, in Fig.~\ref{fig:popcount} with the case of two 18-bit inputs, our method reduces logic depth from 5 to 2 stages and achieves a 52\% reduction in LUT6 usage, thereby improving timing closure and enhancing the overall efficiency of the binary engine.

\subsection{Flexible Orchestrator} \label{sec:orchestrator}
The orchestrator, introduced in Section~\ref{subsec:overall}, receives the current layer configuration from the PS and propagates it downstream—first to the sparse engine, followed by the binary engine. Additionally, it transforms the input layout for the sparse engine, which supports flexible kernel configurations and a data layout derived from tiling along multiple parallel dimensions. This transformation will be elaborated below.

\subsubsection{\textbf{Input Dataflow Orchestration}} \label{subsubsec:push}
According to the parallelism and dataflow outlined in Section~\ref{subsec:parallelism}, the orchestrator's input dataflow is structured as $(C_o / P_{C_o}, T_s \times F_h \times F_w \times C_i / (P_{T_s} \times P_{F_x} \times P_{C_i}), P_{T_s} \times P_{F_x} \times P_{C_i})$, while the output dataflow is defined as $(C_o / P_{C_o}, T_s \times F_h \times F_w \times K_h \times K_w \times C_i /(P_{T_s} \times P_{F_x} \times P_{C_i}), P_{T_s} \times P_{F_x} \times P_{C_i})$. In our design, $P_{F_x}$ performs tiling along the $C_i$ dimension for the input and along the spatial dimension for the output, since tiling by channels improves padding efficiency by replacing per-pixel operations with a single padding step per vectorized channel slice.

To support this dataflow transformation in the push phase (Fig.~\ref{fig:orchestrator}), we employ a banked memory architecture with $P_{F_x}$ banks. Each bank features a bit width of $P_{T_s} \times P_{C_i}$, and uses an interleaved push-addressing mechanism. Initially, the push addresses across the banks are offset by a stride of 1 and incremented by a stride of $P_{F_x}$, allowing $P_{F_x}$ channel slices to be written to sequential addresses. After each full iteration over $C_i / (P_{F_x} \times P_{C_i})$ channel slices, the push address pointer resets, and a cyclic bank-shifting mechanism rotates both the address pointers and the data slices across the banks. This results in an interleaved layout where channel slices of different spatial grids are stored with rotated bank mappings.

Therefore, during the pop phase—when all banks access the same address in each cycle with address increments continuously, and the fetched data slices are reverse-rotated to undo the earlier push-time shift—the layout associated with $P_{F_x}$ transitions from a channel-wise format to spatial tiling, naturally aligning with the target output data organization.

\subsubsection{\textbf{Output Dataflow Orchestration}}
The output dataflow requires structured iteration over the dimensions, starting with $C_i / P_{C_i}$, followed by $K_h \times K_w$. The same pop mechanism described in Section~\ref{subsubsec:push} is employed to retrieve the first channel-in block for the kernel window. Fig.~\ref{fig:orchestrator} illustrates this behavior at the cycle level during its pop phase, specifically for the case where $k_w = 0$ and $c_i = 0 \sim 7$.

To progress through the $K_h \times K_w$ dimension, the memory addressing mechanism must adapt to the kernel window's horizontal sliding. As illustrated in the pop phase of Fig.~\ref{fig:orchestrator}, after each full iteration over $C_i / P_{C_i}$ for a given $k_w$, most bank address pointers reset to their initial state ($c_i = 0$, stored in the base address register). Exceptions are the first $k_w$ banks; their addresses are updated by circularly incrementing from the preceding bank's address (with wrap-around), fetching the new data column necessitated by the kernel's shift. The rotation counter also resets to $k_w$, instead of its initial state, to preserve spatial alignment. By extending this control principle to the remaining outer loops, the system flexibly orchestrates the target output dataflow without any performance stalls.

Unlike Fang et al.'s overlay architecture~\cite{fang2024energy} with its fixed $3 \times 3$ compute grid tailored for $3 \times 3$ kernel operations—resulting in resource underutilization for other kernel sizes such as $1 \times 1$—our orchestrator flexibly supports arbitrary kernel configurations without compromising throughput or resource efficiency. This adaptability stems from a lightweight, configurable register space as shown in Fig.~\ref{fig:orchestrator}. Managed by the PS side of the Zynq SoC, these registers define control parameters (e.g., stride, rotation, layer-specific settings) to direct the dedicated push/pop control logic, while also propagating metadata—such as accumulation loop counts and attention enable flags—to the downstream sparse and binary engines.

\section{Experiments} \label{sec:experiments}

\begin{figure*}
  \centering
  \includegraphics[width=1.0\linewidth]{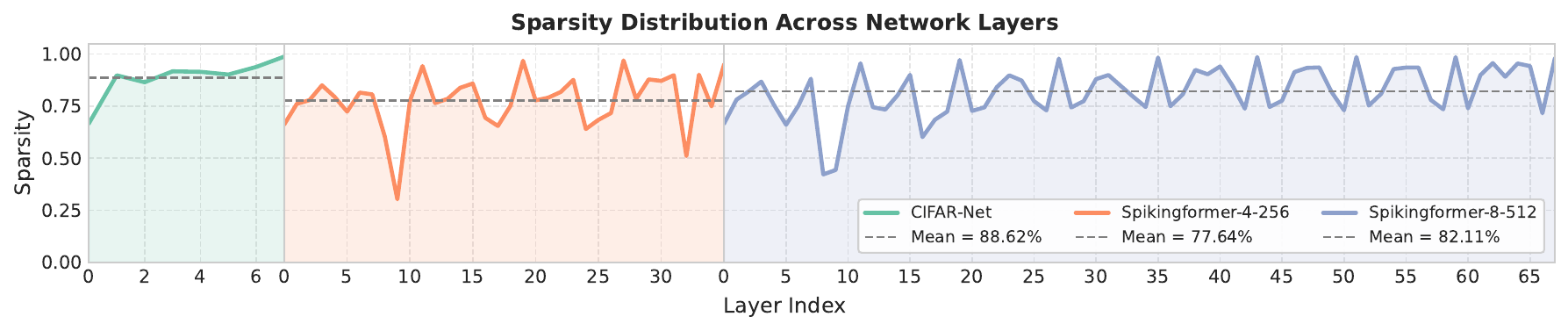}
  \caption{
    Layer-wise sparsity distribution in CIFAR-Net, Spikingformer-4-256, and Spikingformer-8-512.
  }
  \label{fig:sparsity}
\end{figure*}

\begin{figure*}
  \centering
  \includegraphics[width=1.0\linewidth]{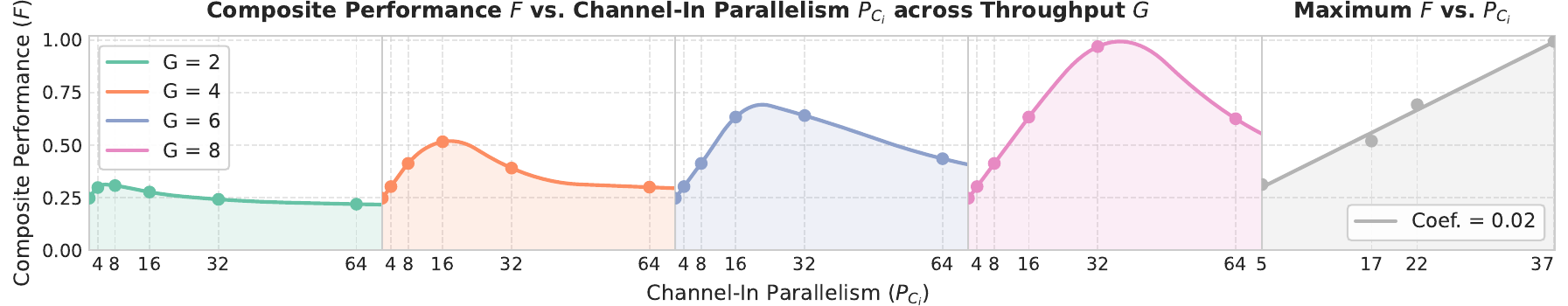}
  \caption{
    Simulation results illustrating the relationship between normalized composite performance ($F$) and channel-in parallelism ($P_{C_i}$) across various throughput ($G$). Left: $F$ versus $P_{C_i}$ for several throughput values ($G$), assuming 75\% sparsity. Right: Maximum achieved $F$ across different $P_{C_i}$.
  }
  \label{fig:channel-in}
\end{figure*}


\subsection{Experimental Setup}
To evaluate FireFly-T's performance, we conduct experiments on several SNN models and datasets: CIFAR-Net (as in FireFly v2~\cite{li2024firefly}) and Spikingformer-4-256 on CIFAR-10, Spikingformer-2-512 on CIFAR-100, and Spikingformer-8-512 on ImageNet. The numerals indicate the number of encoder blocks and the embedding dimension, respectively. All models are trained using the BrainCog framework~\cite{zeng2023braincog} and quantized to 4-bit precision following FireFly-S~\cite{li2024fireflys}.

Hardware deployment and validation are performed on the Xilinx Zynq UltraScale+ FPGA, using the edge-oriented KV260 Vision AI Starter Kit. FireFly-T is developed in SpinalHDL, with Verilog generated via its compiler. These Verilog files are synthesized and implemented through Xilinx Vivado 2024.2. Power consumption estimates are then obtained from the Vivado's implementation reports.

\subsection{Sparse Decoder Configuration} \label{subsec:decoder}
Fig.~\ref{fig:sparsity} presents a layer-wise sparsity analysis for CIFAR-Net, Spikingformer-4-256, and Spikingformer-8-512. The findings reveal that spiking transformers exhibit regular sparsity patterns across the encoder blocks, and all SNN models demonstrate a high and relatively stable sparsity (exceeding 75\%) without requiring additional optimization techniques.

For general analysis, we implement a simulator configured under 75\% spike sparsity. Within this framework, the initial optimization focuses on the configuration of $P_{C_i}$, the input bit-width of the sparse decoder, subject to a fixed throughput $G = P_{W_o} \times M$, representing the maximum processing capacity per grid point in the 3D workload balancer to extract valid weights for non-zero activations each cycle.

Following the aforementioned configuration, we define performance $R = 1 / D$, where $D$ is the overall simulated processing latency. To evaluate the performance-resource trade-off, resource efficiency is given by $\eta = R / B$, with $B = \lambda \times P_{C_i}$ denoting resource utilization and $\lambda$ being a constant scaling factor. Our primary goal is to maximize performance $R$ without overly compromising resource efficiency $\eta$. To capture this balance, a composite performance metric $F$ is defined as:
\begin{equation}
  F = \eta R = \frac{1}{\lambda \times P_{C_i} \times D^2}\\
\end{equation}

The simulation results, presented in Fig.~\ref{fig:channel-in}, first reveal that for a given throughput $G$, the optimal input bit-width $P_{C_i}$ that maximizes the composite performance metric $F$ aligns closely with workload characteristics dictated by spike sparsity. For instance, at $G = 4$, a $P_{C_i}$ value of $G / (1 - 0.75) = 16$ is near-optimal for maximizing $F$. This phenomenon is intuitive: for a fixed $G$, exceeding this optimal $P_{C_i}$ offers no additional throughput gains as PEs become saturated, whereas a $P_{C_i}$ below the optimum leads to PE underutilization, making input bandwidth the bottleneck. Crucially, analysis of the maximum $F$ achieved with these optimal $P_{C_i}$ settings (the final subplot of Fig.~\ref{fig:channel-in}) further reveals a clear linear scaling with $P_{C_i}$. This demonstrates the system architecture's \textbf{scalability of channel-in parallelism} without performance saturation.

Subsequently, the impact of the number of decoder lanes $M$ on performance $R$ is evaluated at fixed throughput $G$, using the aforementioned optimal $P_{C_i}$ (16 for $G = 4$ and 32 for $G = 8$). As shown in Fig.~\ref{fig:performance}A, increasing $P_{W_o}$ (which implies reducing $M$) enhances $R$ via improved load balancing with more workers. While these gains exhibit diminishing returns and introduce resource overhead, we observe that a configuration of $P_{W_o}=2$ (implying $M=2$ for $G=4$ and $M=4$ for $G=8$) notably achieves over 80\% of the peak performance. This result highlights that our architecture can deliver substantial performance through multi-lane sparse decoder even with a reduced worker dimension, achieving a favorable balance between throughput and resource efficiency.

\subsection{Load Balancing Evaluation} \label{subsec:balancing}


\begin{figure*}
  \centering
  \includegraphics[width=1.0\linewidth]{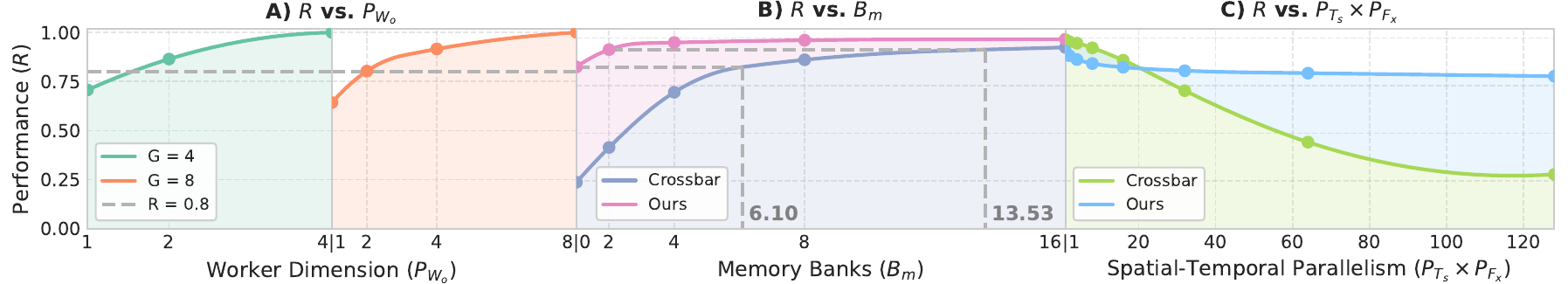}
  \caption{
    Performance analysis across different configurations. A) Relationship between performance $R$ and worker dimension $P_{W_o}$, where $P_{C_i} = 16$ for $G = 4$ and $P_{C_i} = 32$ for $G = 8$. B) Impact of memory bank $B_m$ on performance $R$ with $P_{C_i} = P_{T_s} \times P_{F_x} = 16$. C) Scalability comparison between our method ($B_m = 1$) and the crossbar-based approach~\cite{yin2024loas} ($B_m = 8$) as $P_{T_s} \times P_{F_x}$ increases, with $P_{C_i} = 16$.
  }
  \label{fig:performance}
\end{figure*}

\begin{figure}
  \centering
  \includegraphics[width=1.0\linewidth]{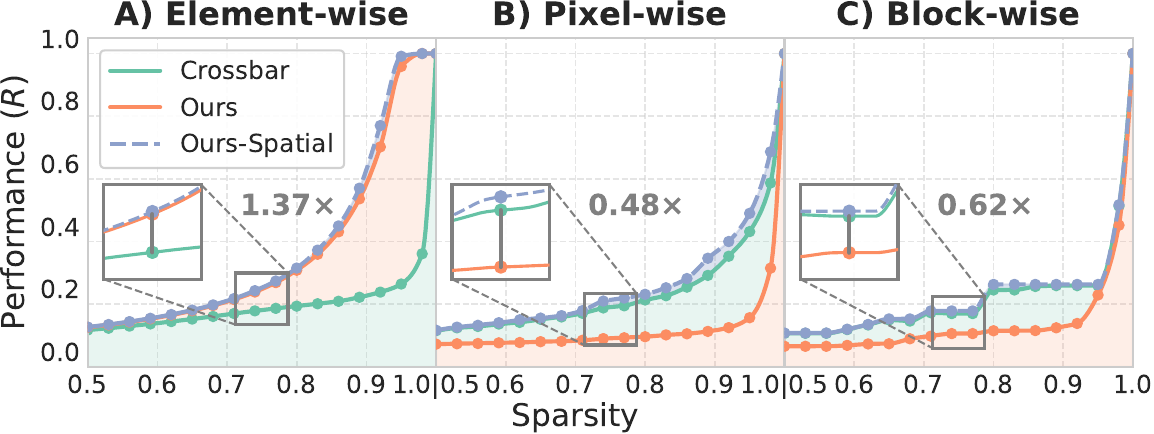}
  \caption{
    Performance sensitivity to various sparsity patterns and levels, where $P_{C_i} = P_{T_s} \times P_{F_x} = 16$, and $B_m=4$ extra banks for the crossbar baseline.
  }
  \label{fig:sensitivity}
\end{figure}

To evaluate our proposed workload balancing mechanism and assess the impact of bank conflicts inherent in crossbar-based method \cite{yin2024loas,parashar2017scnn}, simulations are conducted with a configuration of $P_{C_i} = P_{T_s} \times P_{F_x} = 16$. Notably, our approach utilizes a single-bank design. Therefore, for a fair comparison with consistent total memory bandwidth, the throughput width of our single bank is proportionally scaled as the number of banks in the crossbar method increases.

As presented in Fig.~\ref{fig:performance}B, our approach offers clear advantages. The crossbar-based method's performance substantially deteriorates with insufficient memory banks, primarily due to conflicts caused by data reuse inherent in spatiotemporal parallelism. In contrast, our method exhibits more stable and superior performance. Specifically, with $B_m = 1$ and comparable resource consumption, our approach achieves a $3.48\times$ speedup; conversely, to achieve equivalent performance, the crossbar method would necessitate $6\times$ more memory bandwidth, highlighting our effective workload balancing.

We evaluate scalability by comparing our method ($B_m = 1$) with the crossbar-based approach ($B_m = 8$, chosen to match our initial performance) as the spatiotemporal parallelism $P_{T_s} \times P_{F_x}$ increases, while maintaining $P_{C_i} = 16$. As shown in Fig.~\ref{fig:performance}C, our method achieves better scalability than the crossbar-based approach. Specifically, when $P_{T_s} \times P_{F_x}$ increases from 1 to 128, our method experiences only a 13.17\% performance loss, whereas the crossbar-based method suffers a substantial 70.68\% degradation. These results confirm the enhanced \textbf{scalability of spatiotemporal parallelism} by our proposed workload balancing mechanism, laying the foundation for our multidimensional parallelism design.

\subsection{Sparsity Sensitivity Evaluation} \label{subsec:sparsity}
Fig.~\ref{fig:sensitivity} illustrates the impact of distinct sparsity patterns. While increased sparsity generally improves performance, the sensitivity varies across patterns. At 75\% sparsity, our proposed method outperforms the crossbar baseline by $1.37\times$ under element-wise sparsity, but underperforms for pixel-wise ($0.48\times$) and block-wise ($0.62\times$) patterns.

These observations align with our design expectations. As noted in \textbf{Observation 1} (Section~\ref{subsubsec:balancing}), SNN workloads exhibit relatively uniform distribution across spatio-temporal dimensions at kernel granularity ($K_h \times K_w \times C_i$). Element-wise sparsity preserves this balance, allowing our method to fully exploit the conflict-free mechanism. In contrast, pixel- and block-wise sparsity cause severe spatial load imbalance, which—combined with the orchestrator's shared spatial control logic adopted to reduce hardware overhead—introduces implicit synchronization that bottlenecks performance.

Notably, the crossbar baseline does not enforce this spatial sharing constraint. Under this setting (denoted as Ours-Spatial in Fig.~\ref{fig:sensitivity}), performance remains consistent with our standard implementation for element-wise sparsity, reaches crossbar-level for block-wise sparsity, and achieves $1.12\times$ speedup over crossbar at 75\% sparsity for pixel-wise patterns. The limited improvement under pixel-wise and block-wise patterns stems from the baseline's inherently low conflict rate: parallel PEs either access dense vectors (coalesced accesses) or all-zero vectors (no access), naturally minimizing bank conflicts; however, the crossbar baseline still requires extra banks and corresponding arbiters, incurring higher hardware overhead.

\subsection{Comprehensive Evaluation}
\textbf{Parallelism and Sparsity Support.} Table~\ref{tab:parallelism} compares recent high-performance SNN accelerators regarding their parallelism and sparsity support. A common challenge is that achieving fine-grained sparsity support often introduces dataflow irregularity, which complicates parallel execution and typically leads to load imbalance. To the best of our knowledge, FireFly-T is the first SNN accelerator designed to exploit fine-grained sparsity while maintaining multidimensional parallelism scalability across spatiotemporal, input-channel, and output-channel dimensions, as experimentally validated in Sections~\ref{subsec:decoder} and~\ref{subsec:balancing}. This capability is primarily due to the co-designed sparse engine, which integrates a scalable load-balancing 3D array with high-throughput sparse decoders optimized for fine-grained sparsity exploitation.




\begin{table}[!t]
  \setlength{\tabcolsep}{3.75pt}
  \centering
  \renewcommand\arraystretch{1.25}
  \begin{threeparttable}[b]
  \caption{Parallelism and Sparsity Support of Existing Accelerators}
  \begin{tabular}{cccccc}
  \toprule
  \textbf{Work} & \textbf{Spatial} & \textbf{Temporal} & \textbf{Ch-Out} & \textbf{Ch-In} & \textbf{Sparsity}\\
  \midrule
  TC'23 \cite{aung2023deepfire2} & Yes & No & Yes & Yes & Not Supported \\
  TCAD'24 \cite{li2024firefly} & Yes & Yes & Yes & Yes & Not Supported \\
  MICRO'24 \cite{yin2024loas} & Yes & Yes & No & Yes & Coarse-Grained \\
  JSSC'24 \cite{fang2024energy} & No & Yes & Yes & Yes & Coarse-Grained \\
  TCAS-I'24 \cite{li2024fireflys} & No & No & Yes & Yes & Fine-Grained \\
  \midrule
  \textbf{Ours} & \textbf{Yes} & \textbf{Yes} & \textbf{Yes} & \textbf{Yes} & \textbf{Fine-Grained}\\
  \bottomrule
  \end{tabular}
  \label{tab:parallelism}
  \end{threeparttable}
\end{table}

\begin{table*}
  \setlength{\tabcolsep}{4.325pt}
  \centering
  \renewcommand\arraystretch{1.25}
  \begin{threeparttable}[b]
  \caption{Comparison with Related Work for Multiple Image Classification Tasks.}
  \label{tab:comparison} 
  \begin{tabular}{ccccccccccccc}
  \toprule
  \multirow{2.5}{*}{\makecell{\textbf{Work} \\ \textbf{(Type)}}} & \multicolumn{3}{c}{\textbf{Network}} & \multicolumn{4}{c}{\textbf{Performance}} & \multicolumn{3}{c}{\textbf{Resource}} & \multicolumn{2}{c}{\textbf{Deployment}} \\ \cmidrule(lr){2-4} \cmidrule(lr){5-8} \cmidrule(lr){9-11} \cmidrule(lr){12-13}
  & \textbf{Model} & \textbf{Dataset} & \textbf{Acc.} & \textbf{FPS} & \textbf{GOP/s} & \textbf{GOP/s/W} & \textbf{DSP Eff.} & \textbf{kLUTs} & \textbf{B/URAMs} & \textbf{DSPs} & \textbf{Freq.} & \textbf{Device} \\
    
  \midrule
  \multirow{3}{*}{\makecell{E3NE \cite{gerlinghoff2021e3ne} \\ (D+C)\tnote{1}}} & LeNet5 & MNIST & 99.30 & 3401 & 11 & 3.33 & - & 27 & - & - & 200 & \multirow{3}{*}{xcvu13p} \\
  & AlexNet & CIFAR-10 & 80.60 & 14 & 3 & 0.80 & - & 48 & - & - & 150 &  \\
  & VGG11 & CIFAR-100 & 65.00 & 6 & 11 & 2.24 & - & 88 & - & - & 150 & \\
  \midrule

  \multirow{3}{*}{\makecell{SyncNN \cite{panchapakesan2022syncnn} \\ (S+C)}}  & LeNet-L & MNIST & 99.60 & 1629 & 55 & 139.32\tnote{2} & 0.10 & 224 & - & 554 & 200 & \multirow{3}{*}{xczu9eg} \\
  & NIN & CIFAR-10 & 88.09 & 147 & 52 & 131.93\tnote{2} & - & - & - & - & 200 & \\
  & VGG13 & SVHN & 95.65 & 65 & 29 & 74.24\tnote{2} & - & - & - & - & 200 & \\
  \midrule

  \multirow{3}{*}{\makecell{DeepFire2 \\ \cite{aung2023deepfire2} \\ (D+C)}} & VGG9 & CIFAR-10 & 87.10 & 23000 & 543 & 70.54 & 0.26 & 125 & 511/80 & 2025 & 550 & \multirow{3}{*}{xcvu9p} \\
  & VGG9 & CIFAR-100 & 65.90 & 11600 & 10400 & 518.00 & 3.60 & 183 & 289/452 & 2881 & 500 & \\
  & VGG11 & ImageNet & 40.10 & 1560 & 21100 & 447.00 & 3.90 & 371 & 1757/960 & 5400 & 450 & \\
  \midrule

  \multirow{3}{*}{\makecell{FireFly v2 \cite{li2024firefly} \\ (D+C)}} & CIFAR-Net\tnote{3} & CIFA-10 & 93.60 & 334 & 3443 & 702.74 & 6.73 & 26 & 87/8 & 512 & 250/500\tnote{5} & \multirow{3}{*}{xczu5ev} \\
  & CIFAR-Net\tnote{4} & CIFAR-100 & 74.70 & 222 & 3625 & 739.81 & 7.08 & 26 & 87/8 & 512 & 250/500\tnote{5} & \\
  & SEW-ResNet34 & ImageNet & 62.4 & 40 & 3103 & 633.33 & 6.06 & 26 & 87/8 & 512 & 250/500\tnote{5} &  \\
  \midrule

  \multirow{2}{*}{\makecell{HeatViT \cite{dong2023heatvit} \\ (S+T)}} & DeiT-T & ImageNet & 72.20 & 183 & 440 & 46.82 & 0.22 & 137 & 355/- & 1968 & 150 & \multirow{2}{*}{xczu9eg} \\
  & DeiT-B & ImageNet & 80.10 & 36 & 12500 & 110.13 & 6.05 & 161 & 589/- & 2066 & 150 & \\
  \midrule

  \multirow{2}{*}{\makecell{SSR \cite{zhuang2024ssr} \\ (D+T)}} & DeiT-T & ImageNet & 72.20 & 4540 & 10900 & 246.15 & 6.06 & 620 & 624/104 & 1797 & 230/1000\tnote{6} & \multirow{2}{*}{xcvc1902} \\
  & LV-ViT-T & ImageNet & 79.10 & 2630 & 8210 & 181.74 & - & - & - & - & 230/1000\tnote{6} & \\
  \midrule

  \multirow{3}{*}{\makecell{SpikeTA \cite{gao2025advancing} \\ (D+T)}} & Spikingformer\tnote{7} & CIFAR-10 & 95.60 & - & 28980 & 408.57 & 3.99 & 503 & 1366/576 & 7249 & 450 & \multirow{3}{*}{xcu280} \\
  & Spikingformer\tnote{7} & CIFAR-100 & 78.40 & - & 28980 & 405.60 & 3.99 & 503 & 1366/576 & 7249 & 450 & \\
  & Spikingformer\tnote{8} & ImageNet & 74.56 & - & 28990 & 403.99 & 4.04 & 501 & 1572/594 & 7168 & 450 & \\
  \midrule

  \multirow{4}{*}{\makecell{\textbf{FireFly-T} \\ \textbf{(S+C+T)}}} & \textbf{CIFAR-Net\tnote{3}} & \textbf{CIFAR-10} & \textbf{93.60} & \textbf{354} & \textbf{3630} & \textbf{978.61} & \textbf{28.35} & \textbf{24} & \textbf{74/0} & \textbf{128} & \textbf{300} & \multirow{4}{*}{\textbf{xczu5ev}} \\
  & \textbf{Spikingformer\tnote{9}} & \textbf{CIFAR-10} & \textbf{94.45} & \textbf{907} & \textbf{3029} & \textbf{696.64} & \textbf{9.96} & \textbf{46} & \textbf{100/0} & \textbf{304} & \textbf{300} & \\
  & \textbf{Spikingformer\tnote{7}} & \textbf{CIFAR-100} & \textbf{78.38} & \textbf{630} & \textbf{3317} & \textbf{762.87} & \textbf{10.91} & \textbf{46} & \textbf{100/0} & \textbf{304} & \textbf{300} & \\
  & \textbf{Spikingformer\tnote{8}} & \textbf{ImageNet} & \textbf{71.55} & \textbf{50} & \textbf{3397} & \textbf{781.13} & \textbf{11.11} & \textbf{46} & \textbf{100/0} & \textbf{304} & \textbf{300} & \\
  \bottomrule
  \end{tabular}
  \begin{tablenotes}
  \item[1] D: Dense, S: Sparse, C: Convolution, T: Transformer
  \item[2] SyncNN calculates power usage by measuring the active power consumption of the FPGA board (24.5 W) and subtracting the static power measured when the board is idle (24.1 W). In contrast, most existing research typically acquires power metrics directly from the Vivado report.
  \item[3] 3x32x32-32c3-256c3-256c3-mp2-256c3-256c3-256c3-mp2-512c3-mp2-1024c3-ap-10
  \item[4] 3x32x32-64c3-256c3-256c3-mp2-256c3-512c3-512c3-mp2-512c3-mp2-1024c3-ap-10
  \item[5] The base clock frequency of FireFly v2 is 250 MHz, with a DSP double data rate clock of 500 MHz.
  \item[6] SSR is implemented on the AMD ACAP VCK190 board, with PL operating at 230 MHz and the AI Engine (AIE) running at 1 GHz.
  \item[7,8,9] Entries 7, 8, and 9 correspond to Spikingformer-2-512, Spikingformer-8-512, and Spikingformer-4-256, respectively.
\end{tablenotes}
\end{threeparttable}
\end{table*}

\textbf{Performance and Energy Efficiency.} Based on the previous experimental results, we configure the parallelism of FireFly-T as $(P_{T_s} \times P_{F_x}, P_{C_i}, P_{C_o}) = (8, 16, 64)$, aligning with the theoretical peak GOP/s performance of our prior dense accelerator, FireFly v2~\cite{li2024firefly}, on the same hardware platform. The throughput $G$ is set to 2 for CIFAR-Net and 4 for spiking transformer, respectively. These settings are adopted in the comparative experiments presented in Table~\ref{tab:comparison}.

Specifically, E3NE \cite{gerlinghoff2021e3ne} and SyncNN \cite{panchapakesan2022syncnn} represent early-stage SNN accelerators. E3NE employs an instruction-based end-to-end framework but lacks sparsity exploitation, leading to comparatively low throughput (11.33 GOP/s on MNIST). SyncNN utilizes mixed-precision quantization with a spatial architecture, achieving higher energy efficiency compared to E3NE (3.33 vs. 139.32 GOP/s/W on MNIST), but demands relatively high LUT and DSP consumption.

Recent efforts such as DeepFire2~\cite{aung2023deepfire2} and FireFly v2~\cite{li2024firefly} have demonstrated high performance: DeepFire2 achieves notable throughput (1.56 kFPS and 21.10 TOP/s on ImageNet), while FireFly v2 delivers higher energy efficiency (633.33 vs. 447.00 GOP/s/W on ImageNet). However, DeepFire2's spatial architecture demands per-layer logic instantiation, consuming substantial resources (1757/960 B/URAMs and 5400 DSPs), and both works lack sparsity support and are limited to spiking convolutional networks, constraining network accuracy. FireFly-T overcomes these limitations, achieving $1.47\times$--$13.87\times$ and $1.03\times$--$1.39\times$ higher energy efficiency than DeepFire2 and FireFly v2, respectively, across the CIFAR-10, CIFAR-100, and ImageNet datasets. By exploiting fine-grained sparsity, FireFly-T improves DSP efficiency (GOP/s/DSP) by $4.21\times$ over FireFly v2 on CIFAR-Net. Even with the more complex Spikingformer-8-512 architecture (71.55\% accuracy), FireFly-T consumes $17.76\times$ fewer DSPs and although it exhibits a $6.21\times$ lower GOP/s than DeepFire2's VGG11 baseline (40.10\% accuracy), it still delivers a $2.85\times$ improvement in DSP efficiency. Overall, compared to DeepFire2 and FireFly v2, FireFly-T demonstrates superior energy and DSP efficiency, with better accuracy and fewer resources.

Several prior efforts have targeted transformer acceleration. HeatViT~\cite{dong2023heatvit} introduced image-adaptive token pruning, achieving 1.25 TOP/s throughput. SSR~\cite{zhuang2024ssr} improves energy efficiency over HeatViT, from 46.82 GOP/s/W to 246.15 GOP/s/W. Despite their notable network performance, both architectures demand substantial hardware resources (1968 DSPs for HeatViT; 1797 DSPs and 394 AIEs for SSR), due to the computationally intensive dense MAC operations, limiting their energy efficiency. SpikeTA~\cite{gao2025advancing} supports spiking transformer with DSP-efficient addition trees and depth-aware buffer management, achieving enhanced energy efficiency of 403.99 GOP/s/W. However, its scope is confined to the encoder block, and the spatial architecture without sparsity support limits deployment to the costly U280 device. In comparison, FireFly-T achieves $1.88\times$ to $16.68\times$ better energy efficiency and $1.83\times$ to $50.50\times$ higher DSP efficiency than these transformer-enabled accelerators. Utilizing an overlay architecture, FireFly-T flexibly supports various network topologies with an order of magnitude fewer resources, enabling deployment on resource-constrained edge devices.



\begin{table}
  \setlength{\tabcolsep}{3.75pt}
  \centering
  \renewcommand\arraystretch{1.25}
  \caption{
    Breakdown of Resource Usage for FireFly-T with $G=4$.
  }
  \label{tab:resource}
  \begin{tabular}{ccccc}
    \toprule
    \textbf{Component} & \textbf{kLUTs} & \textbf{kREGs} & \textbf{BRAMs} & \textbf{DSPs} \\
    \midrule
    Sparse Engine & 40 (34.0\%) & 109 (46.8\%) & 72 (50.0\%) & 288 (23.1\%) \\
    Binary Engine & 5 (4.1\%) & 9 (3.7\%) & 24 (16.7\%) & 16 (1.3\%) \\
    Orchestrator & 1 (1.2\%) & 1 (0.5\%) & 4 (2.7\%) & 0 (0.0\%) \\
    \midrule
    \textbf{Total} & \textbf{46 (39.5\%)} & \textbf{117 (49.8\%)} & \textbf{100 (69.5\%)} & \textbf{304 (24.4\%)} \\
    \bottomrule
  \end{tabular}
\end{table}

\textbf{Resource Breakdown.} Table~\ref{tab:resource} provides a detailed breakdown of resource utilization within FireFly-T where $G$ is set to 4. As shown, the binary engine consumes only 11.98\% of the LUTs used by the sparse engine, primarily due to its 1-bit operand design—enabled by the spiking attention mechanism—and an optimized implementation of the AND-PopCount logic. Additionally, the binary engine utilizes just 6.10\% of the DSP resources compared to the sparse engine, consistent with the analysis in Section~\ref{subsec:pipeline}. These results highlight the resource efficiency of the binary engine.

Table~\ref{tab:lut} presents the LUT breakdown within the sparse engine, allowing an assessment of resource overhead from fine-grained sparsity support. The sparse decoder, dedicated to input bitmap decoding, utilizes only 5.73\% ($G = 2$) and 3.62\% ($G = 4$) of the engine's LUTs. This efficiency is a direct result of our input reuse strategy, which permits broadcasting decoded indices across output channels. In contrast, the load-balancing array commands the majority of LUT resources (75.86\% for $G = 2$ and 84.20\% for $G = 4$). This is expected, as it forms the engine's core computational logic for workload distribution and valid weight extraction within PEs.
\begin{table}
  \setlength{\tabcolsep}{4pt}
  \centering
  \renewcommand\arraystretch{1.25}
  \caption{
    Breakdown of LUT Usage for FireFly-T's Sparse Engine.
  }
  \label{tab:lut}
  \begin{tabular}{ccccc}
    \toprule
    \textbf{Config} & \textbf{Decoder} & \textbf{Balancer} & \textbf{Neuron} & \textbf{Others} \\
    \midrule
    $G = 2$ & 1306 (5.73\%) & 17280 (75.86\%) & 2138 (9.39\%) & 2055 (9.02\%) \\
    $G = 4$ & 1442 (3.62\%) & 33536 (84.20\%) & 2293 (5.76\%) & 2557 (6.42\%) \\
    \bottomrule
  \end{tabular}
\end{table}

Despite the additional resource overhead, the trade-off turns out to be highly beneficial: enabling sparsity support substantially reduces DSP usage, achieving a fractional reduction of $1-G/P_{C_i}$. Under our parallelism setup, this corresponds to DSP savings of 896 ($G=2$) and 768 ($G=4$), compared to 1024 4-lane DSPs required by the dense method. According to the conversion factor from \cite{li2023firefly} (1 DSP $\approx$ 86 LUTs), these reductions are equivalent to 77k and 66k LUTs, respectively. In contrast, the combined logic overhead from the decoder and load-balancing is significantly lower—18k LUTs ($G=2$) and 34k LUTs ($G=4$), highlighting the efficiency and practicality of the proposed design for sparse acceleration.

\section{Related Work} \label{sec:related}
Recent work has focused on hardware accelerators for spiking convolutional networks~\cite{li2023firefly,li2024firefly, li2024fireflys, aung2023deepfire2}. FireFly v2~\cite{li2024firefly} proposes an overlay architecture with spatiotemporal dataflow and double data rate clocking on DSPs~\cite{li2024revealing}. FireFly-S~\cite{li2024fireflys} presents a software-hardware co-design leveraging dual-side sparsity. DeepFire2~\cite{aung2023deepfire2} introduces a spatial architecture employing split-kernel mapping to efficiently scale up SNN processing. Despite these advancements, existing designs either fail to effectively leverage sparsity in overlay architectures or perform poorly on large-scale datasets such as ImageNet, making them unsuitable for adaptation to spiking transformers.

Several accelerators have been proposed for non-spiking transformer models~\cite{wang2022via, dong2023heatvit, zhuang2024ssr}. ViA~\cite{wang2022via} customizes the hardware design by employing a half-layer mapping strategy tailored for ViT. HeatViT~\cite{dong2023heatvit} introduces a hardware-efficient, image-adaptive token pruning framework to optimize performance. SSR~\cite{zhuang2024ssr} presents a sequential spatial hybrid architecture that explores the trade-off between latency and throughput. Although these non-spiking transformer accelerators achieve high prediction accuracy, they typically lack energy efficiency and demand high-end hardware resources.

In the emerging field of spiking transformer acceleration, SpikeTA~\cite{gao2025advancing} introduces DSP-efficient addition trees and depth-aware buffer management techniques. However, its spatial architecture limits resource reuse and lacks mechanisms for exploiting sparsity. Similarly, Fang et al.~\cite{fang2024energy} propose an overlay architecture based on a fixed $3 \times 3$ compute grid, which suffers from performance degradation under diverse kernel configurations and fails to leverage fine-grained sparsity to further improve energy efficiency.

\section{Conclusion} \label{sec:conclusion}
FireFly-T substantially advances our prior designs, FireFly v2 and FireFly-S, distinguished by its support for spiking transformers, high-throughput fine-grained sparsity exploitation, and multidimensional parallelism, all enabling efficient end-to-end SNN deployment. This is achieved through a dual-engine overlay architecture: the sparse engine employs multi-lane sparse decoders and bank-conflict-free workload balancers to effectively exploit sparse inputs, with enhanced scalability for multidimensional parallelism; the binary engine utilizes implicit dataflow manipulation and LUT6-Optimized AND-PopCount logic, enabling resource-efficient spiking attention support. Additionally, an orchestrator flexibly manages dataflow, ensuring adaptability to diverse network topologies.

Experimental validation confirms FireFly-T's superior energy and DSP efficiency across a range of datasets and SNN architectures. Relying on commercial FPGA edge devices, FireFly-T offers a practical and accessible hardware solution. In future work, we aim to advance SNN hardware that effectively supports the rapid evolution of SNN algorithms and enables improved software-hardware co-design.



\bibliography{main}
\bibliographystyle{IEEEtran}

\vspace{-50pt}
\begin{IEEEbiography}
  [{\includegraphics[width=1in,height=1.25in,clip,keepaspectratio]{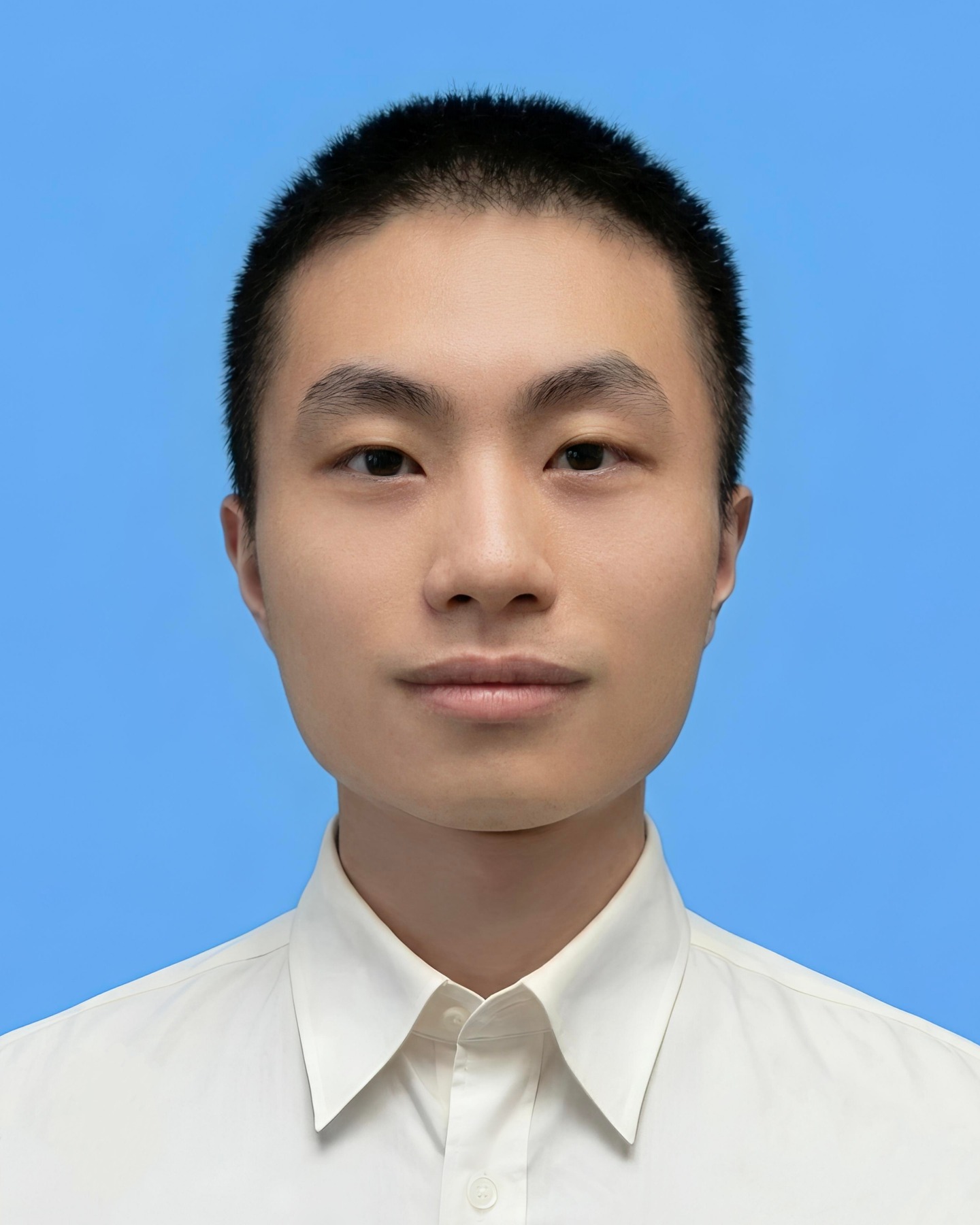}}]
  {Tenglong Li} received his bachelor degree from Sun Yat-sen University in Guangzhou, Guangdong, China in 2023. He is now a PhD candidate in the Brain-inspired Cognitive AI Lab, at the Institute of Automation, Chinese Academy of Sciences, under the supervision of Prof. Qian Zhang and Prof. Yi Zeng. His research focuses on hardware acceleration of brain-inspired algorithms, domain-specific architecture and FPGA system design.
\end{IEEEbiography}
\vspace{-50pt}
\begin{IEEEbiography}
  [{\includegraphics[width=1in,height=1.25in,clip,keepaspectratio]{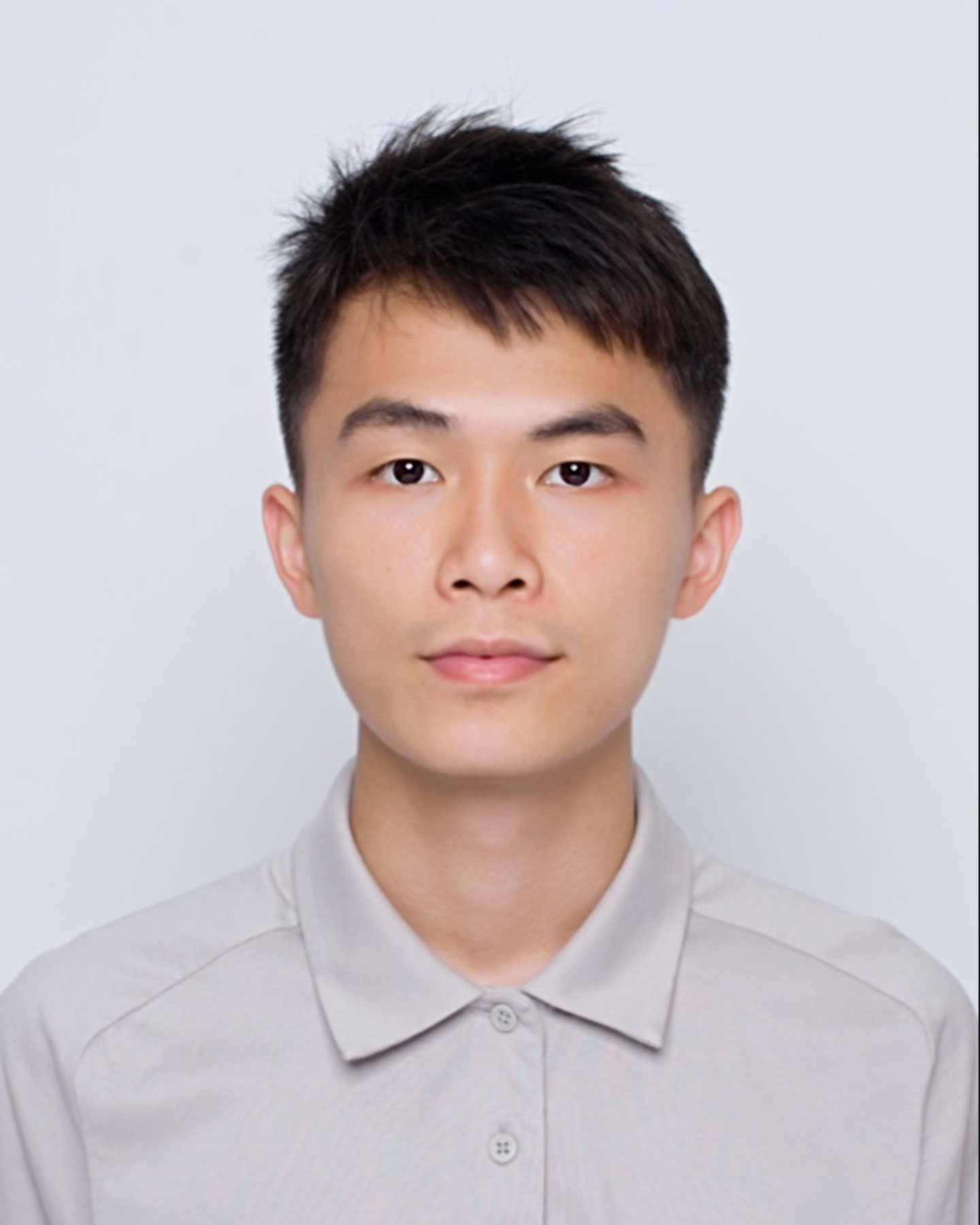}}]
  {Jindong Li} received his bachelor degree from Sun Yat-sen University in Guangzhou, Guangdong, China in 2022. He is now a PhD candidate in the Brain-inspired Cognitive AI Lab, at the Institute of Automation, Chinese Academy of Sciences, under the supervision of Prof. Qian Zhang and Prof. Yi Zeng. His research focuses on hardware acceleration of brain-inspired algorithms, domain-specific architecture and FPGA system design.
\end{IEEEbiography}
\vspace{-50pt}
\begin{IEEEbiography}
  [{\includegraphics[width=1in,height=1.25in,clip,keepaspectratio]{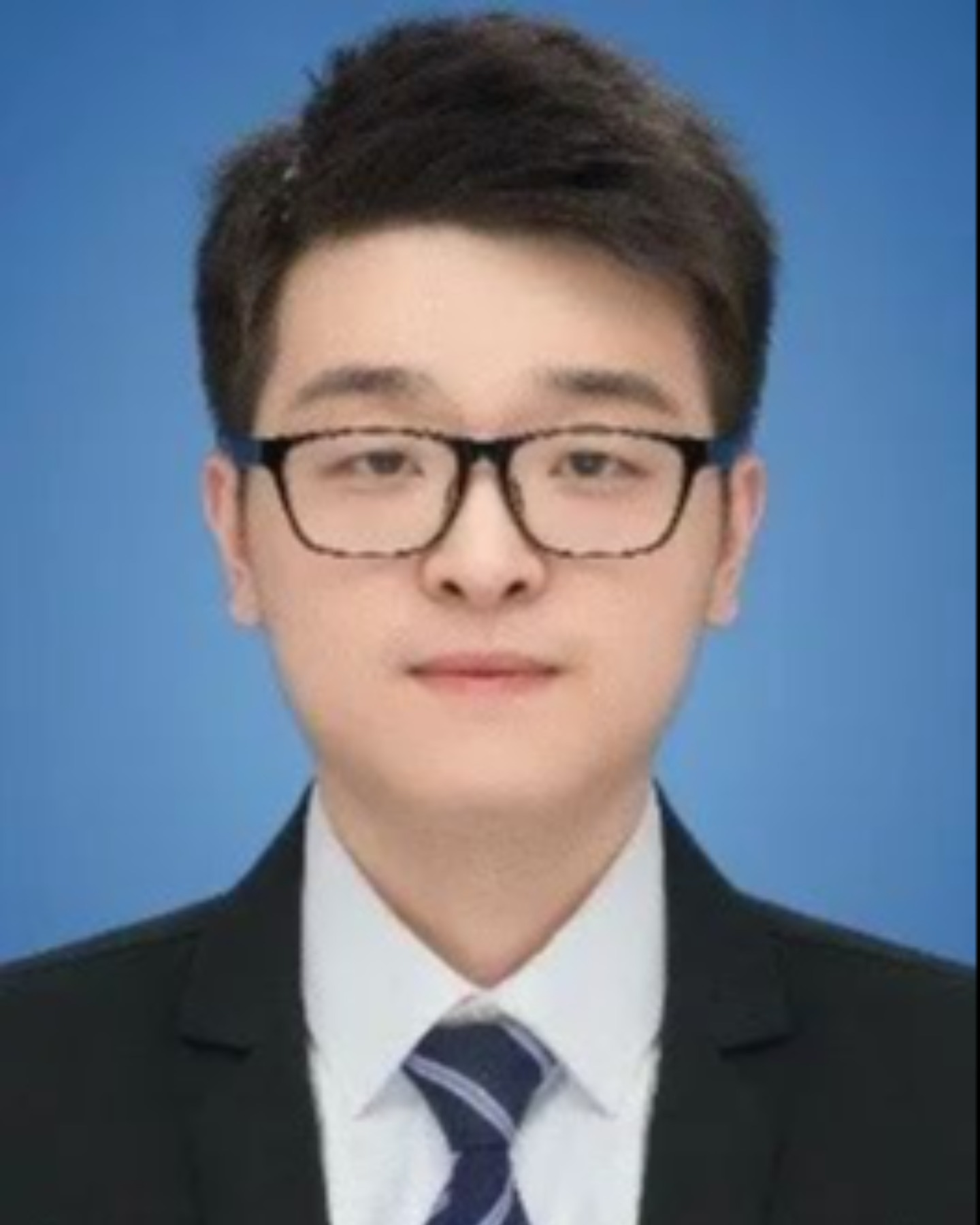}}]
  {Guobin Shen} received his bachelor degree from Sun Yat-sen University in Guangzhou, Guangdong, China in 2021. He is now a PhD candidate in the Brain-inspired Cognitive AI Lab, at the Institute of Automation, Chinese Academy of Sciences, under the supervision of Prof. Yi Zeng. His research focuses on biologically-inspired learning algorithms and spiking neural network architecture design and training strategies.
\end{IEEEbiography}
\vspace{-50pt}
\begin{IEEEbiography}
  [{\includegraphics[width=1in,height=1.25in,clip,keepaspectratio]{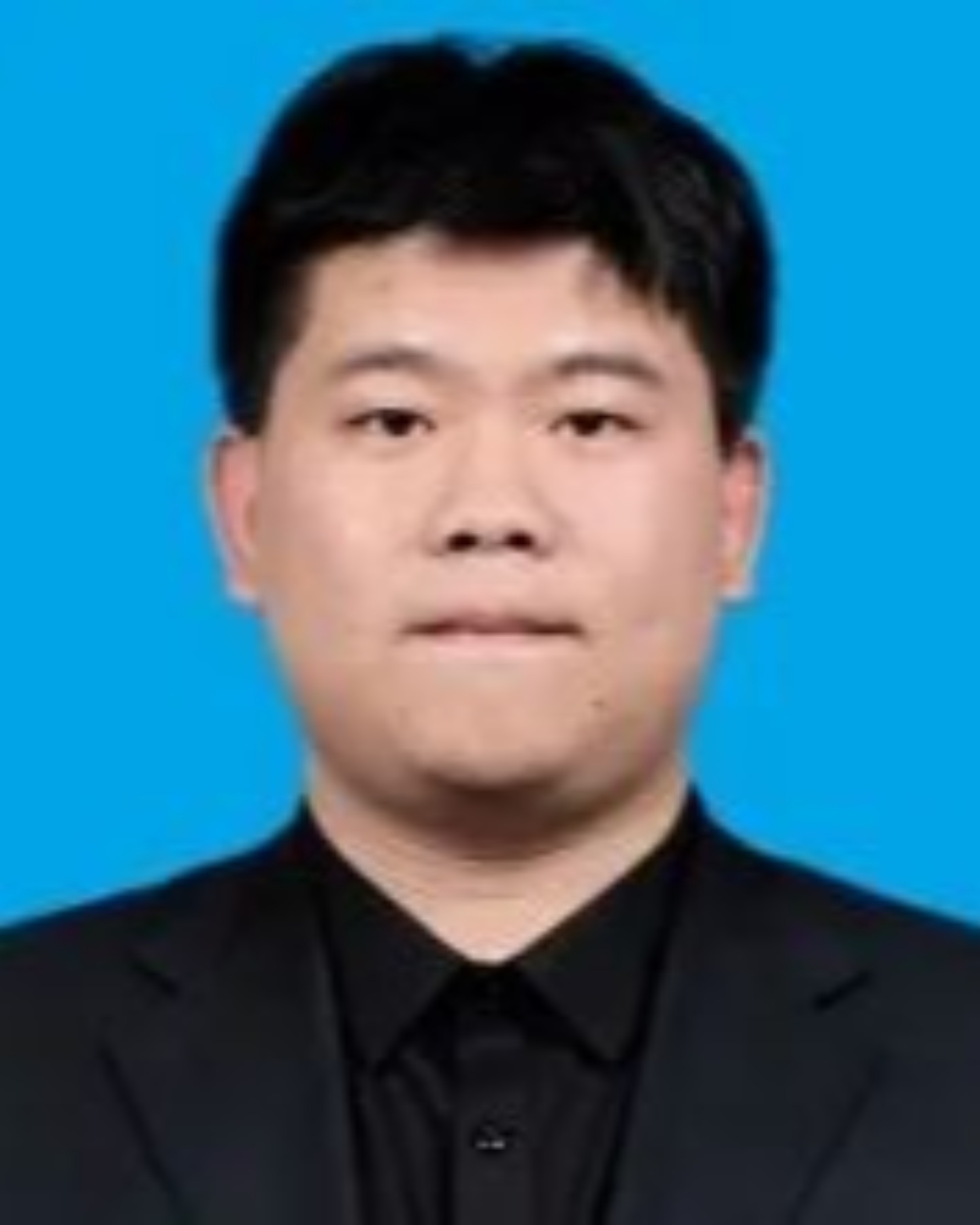}}]
  {Dongcheng Zhao} received the bachelor degree from XiDian University, Xi'an, Shaanxi, China, in 2016 and Ph.D degree from University of Chinese Academy of Sciences, Beijing, China, in 2021. He is currently an assistant professor in the Brain-inspired Cognitive AI Lab, Institute of Automation, Chinese Academy of Sciences, China. His current research interests include learning algorithms in spiking neural networks, thalamus-cortex interaction, visual object tracking, etc.
\end{IEEEbiography}
\vspace{-57pt}
\begin{IEEEbiography}
  [{\includegraphics[width=1in,height=1.25in,clip,keepaspectratio]{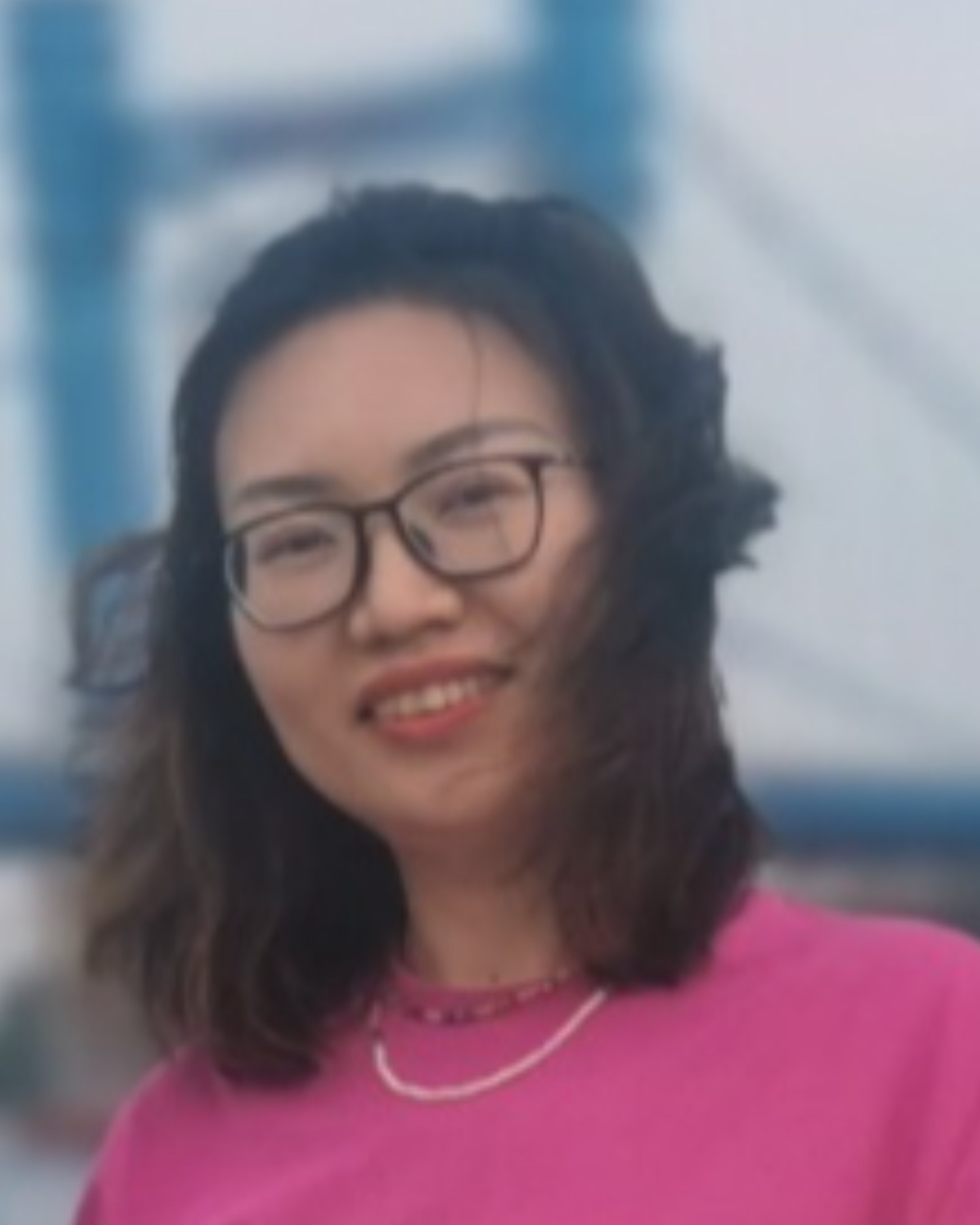}}]
  {Qian Zhang} obtained her bachelor degree in 2009 and Ph.D degree in 2014 from Xidian University, China. She is currently an associate professor and director in the Brain-inspired Cognitive AI Lab, Institute of Automation, Chinese Academy of Sciences, China. Her research interests include brain simulation and brain-inspired cognitive computing modeling, especially working memory modeling and simulation of brain rhythms at different levels of consciousness.
\end{IEEEbiography}
\vspace{-57pt}
\begin{IEEEbiography}
  [{\includegraphics[width=1in,height=1.25in,clip,keepaspectratio]{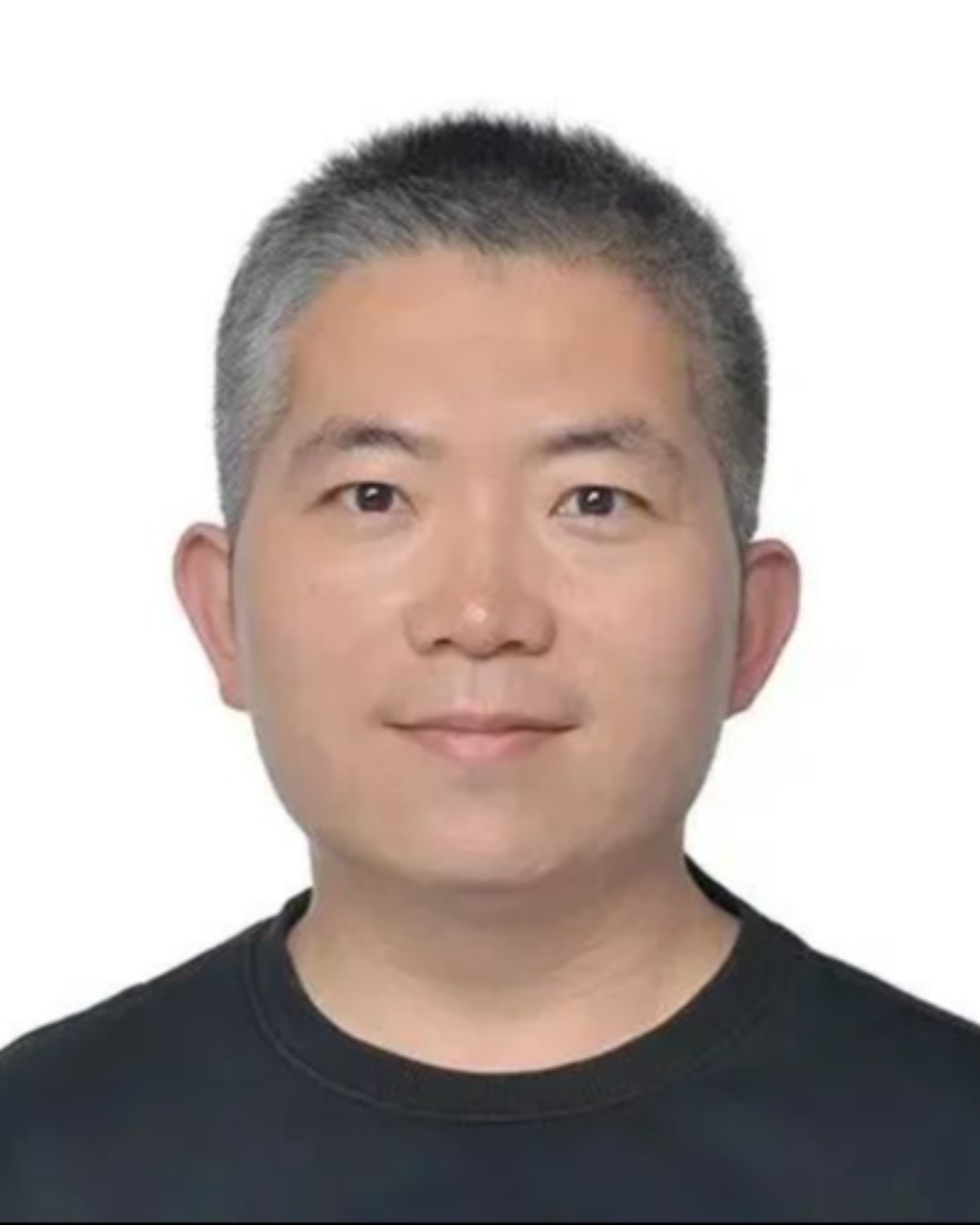}}]
  {Yi Zeng} obtained his bachelor degree in 2004 and Ph.D degree in 2010 from Beijing University of Technology, China. He is currently a professor and director in the Brain-inspired Cognitive AI Lab, Institute of Automation, Chinese Academy of Sciences, China. He is a principal investigator in the Center for Excellence in Brain Science and Intelligence Technology, Chinese Academy of Sciences, China, and a Professor in the School of Future Technology, and School of Humanities, University of Chinese Academy of Sciences, China. His research interests include brain-inspired artificial intelligence, brain-inspired cognitive robotics, ethics and governance of Artificial Intelligence, etc.
\end{IEEEbiography}
\vspace{-17.9pt}


 





\end{document}